\documentclass[12pt]{article}
\usepackage{amsmath}
\usepackage{times}
\usepackage{graphicx}
\usepackage{color}
\usepackage{multirow}
\usepackage[section]{placeins}
\usepackage[numbers]{natbib}
\usepackage{rotating}
\usepackage{bbm}
\usepackage{latexsym}
\usepackage{float}

\newcommand{\captionfonts}{\normalsize}

\makeatletter  
\long\def\@makecaption#1#2{%
  \vskip\abovecaptionskip
  \sbox\@tempboxa{{\captionfonts #1: #2}}%
  \ifdim \wd\@tempboxa >\hsize
    {\captionfonts #1: #2\par}
  \else
    \hbox to\hsize{\hfil\box\@tempboxa\hfil}%
  \fi
  \vskip\belowcaptionskip}
\makeatother   

\begin{document}
\title{Dynamics of Adaptive Continuous Attractor Neural Networks}
\author{}
\date{}
\maketitle

{\bf \large Yujun Li$^{\displaystyle 1}$, Tianhao Chu$^{\displaystyle 3}$, Si Wu$^{\displaystyle 2, \displaystyle 3}$}\\
{$^{\displaystyle 1}$Yuanpei College, Peking University, China}\\
{$^{\displaystyle 2}$School of Psychological and Cognitive Sciences,Peking University, China}\\
{$^{\displaystyle 3}$Academy for Advanced Interdisciplinary Studies, Peking University, China}\\
%


\thispagestyle{empty}
\markboth{}{NC instructions}
\ \vspace{-0mm}\\

\begin{abstract}
Attractor neural networks consider that neural information is stored as stationary states of a dynamical system formed by a large number of interconnected neurons. The attractor property empowers a neural system to encode information robustly, but it also incurs the difficulty of rapid update of network states, which can impair information update and search in the brain. To overcome this difficulty, a  solution is to include adaptation in the attractor network dynamics, whereby the adaptation serves as a slow negative feedback mechanism to destabilize which are otherwise permanently stable states. In such a way, the neural system can, on one hand, represent information reliably using attractor states, and on the other hand, perform computations wherever rapid state updating is involved.   
Previous studies have shown that continuous attractor neural networks with adaptation (A-CANNs) exhibits rich dynamical behaviors accounting for various brain functions. In this paper, we present a comprehensive review of the rich diverse dynamics of A-CANNs. Moreover, we provide a unified mathematical framework to understand these different dynamical behaviors, and briefly discuss about their  biological implications.
\end{abstract}
 
\section{Introduction}
\begin{figure}[H]
\begin{center}
\includegraphics[width=15 cm]{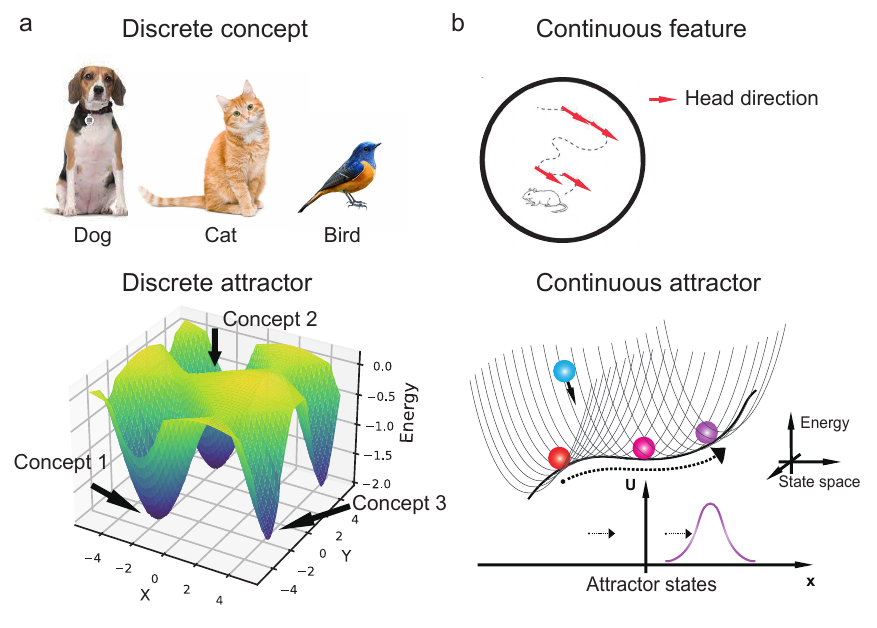}
\end{center}
\caption{
\small
Discrete and continuous attractor network model. 
\textbf{a}. An illustration of discrete concept and discrete attractors. Each attractor represents a single concept or item, such as dog, cat and bird.
\textbf{b}. An illustration of continuous feature and continuous attractor. Head direction cells represent the continuous angular orientations of the rodent's head movement during exploration in a new environment. The stationary states of a CANN form a subspace in which the network states are neutrally stable. The subspace is illustrated as a canyon in the energy space. The movement of the network state along the canyon corresponds to the position shift of the bump.
}
\label{fig0}
\end{figure}

Attractor networks are a class of neural network models which consider that neural information is encoded as stationary states of a dynamical system formed by a large number of interconnected neurons. An early influential attractor network model is the Hopfield network~\cite{hopfield1982neural}, which considers that representations of information (neuronal activity patterns) are stored as local minimums of an energy function of the network dynamics. The Hopfield model typically considers that the stored information representations 
are random patterns, which leads to discrete attractors in the state space (see Fig.\ref{fig0}a). These discrete attractors ensure that once the initial state of the network falls into the basin of an attractor, the network state will robustly converge to it, achieving the so-called associative memory. 

Although the Hopfield model can explain associative memory for independent information representations, it falls to account for situations when representations are highly correlated or the information takes continuous values. In such a case, a new type of attractor network model called continuous attractor neural networks (CANNs) was developed~\cite{amari1977dynamics, ermentrout1998neural, bressloff2011spatiotemporal}, in which the network attractors form a continuous subspace to represent the stored information in a continuous manner (see Fig.\ref{fig0}b), such as the orientation, head-direction, or spatial-location of an object~\cite{ben1995theory, battaglia1998attractor, doboli2000latent}.
A key property of a CANN is that the network is neutrally stable on the attractor space, which allows the neural system to track the movement of an object smoothly by updating the corresponding neural representations continuously, such as for tracking head rotation or performing path integration~\cite{conklin2005controlled, fuhs2006spin}.  

Strictly speaking, both the Hopfield network and CANNs are idealised mathematical models, which correspond to two extreme situations, respectively, i.e., when neural information representations are independent to each other or constitute a continuous spectrum. In reality, neural representations may have a correlation structure in between the two ends, and consequently, the structure of the attractor space of the neural network has a form in between those of the Hopfield network and CANNs. Nevertheless, through studying the simplified Hopfield network and CANNs, it gives us insight into understanding how general attractor networks contribute to brain functions. 

The key property of an attractor network is that it holds information representations as stationary states of the network, which endows the neural system with the capacity of representing information reliably and retrieving information robustly. In practice, however, the neural system needs not only to represent information reliably, but also to process it efficiently. For instance, during a memory recall task, the neural system often needs to search memory candidates over a large state space by updating its current state rapidly. This requires the neural representation to have sufficient mobility, which seems to contradict with the demand of maintaining stability of information representation. To overcome this dilemma, a solution is to introduce an additional mechanism to destabilize attractors to increase the mobility of the network state, and such a mechanism can be neural adaptation.  

Neural adaptation refers to a general phenomenon that a neural system exploits negative feedback to suppress its responses when the activity level is high. Neural adaptation can occur either at the single neuron level via spike frequency adaptation (SFA) ~\cite{bressloff2011spatiotemporal, gutkin2014spike} or at the synapse level via short-term plasticity (STP) ~\cite{abbott1997synaptic, tsodyks1998neural, markram1998differential}, both widely existing in neural systems. 
The time scale of neural adaptation is typically much slower than that of single neuron and synapse dynamics. Thus, by incorporating adaptation in an attractor network, it, on one hand, allows the network to hold an active state in a short time scale to encode information reliably, and on the other hand, enables the network to update states in a long time scale easily.    

In previous works, the authors and their collaborates have studied the model of CANNs with adaptation~\cite{mi2014spike, fung2012dynamical}, referred to as adaptive CANNs (A-CANNs) hereafter. They showed that an A-CANN exhibits rich dynamical behaviors not shared by a CANN without adaptation, and these rich dynamics contribute to various brain functions. In this paper, we provide a comprehensive review of these studies. In particular,
we present new theoretical analyses which link the different dynamical behaviors reported in previous publications, and hence provide a unified mathematical framework to understand these different results. 
Limited by space, our focus is on presenting the mathematical analyses of the rich dynamics of A-CANNs, and their biological implications are only briefly discussed.

The organization of the paper is as follows. In Sec.2, we first introduce a model of A-CANN which adopts SFA as the adaptation mechanism, and then present a projection method to simplify the dynamics of the A-CANN, which is crucial for theoretical analysis. In Sec.3, we study the spontaneous dynamics of the A-CANN, including the static bump state when the adaptation is weak and the travelling wave state when the adaptation strength is large enough. In Sec.4, we study the tracking dynamics of the A-CANN when the network receives an external moving input. We show that depending on the relative strengths of adaptation and external input, the network exhibits three different states, which are smoothing tracking, oscillatory tracking, and travelling wave. The biological implications of these states are discussed. In Sec.5, we study the stochastic dynamics of the A-CANN when the adaptation is noisy. We show that when the noisy adaptation strength is around the boundary of travelling wave, the spontaneous dynamics of the network state exhibits the characteristic of Lévy flight. In Sec.6, overall conclusions and discussions of this work are given. Finally, Appendixes present the detailed mathematical analyses of the network model.

\section{Adaptive Continuous Attractor Neural Networks}

\begin{figure}[h]
\begin{center}
\includegraphics[width=15 cm]{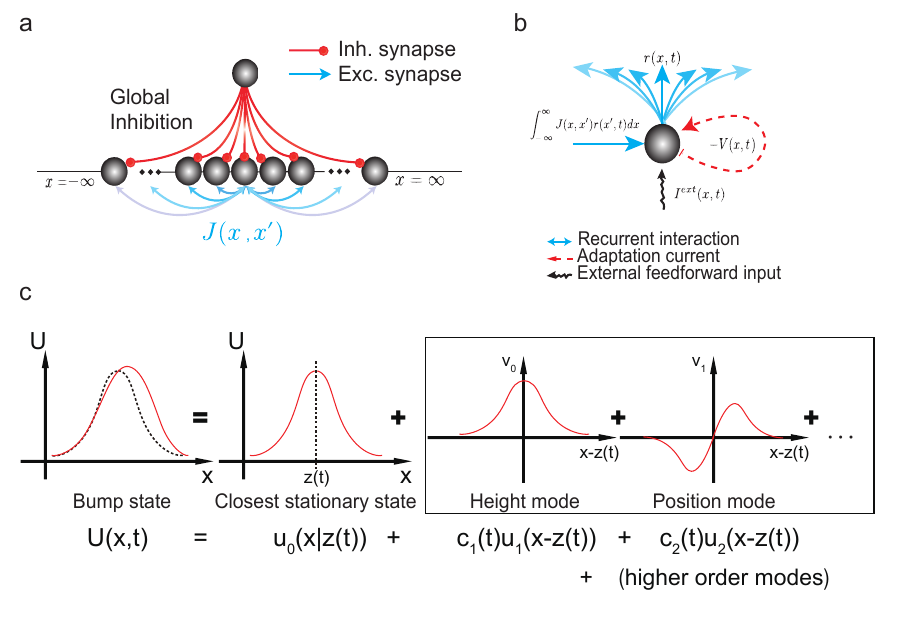}
\end{center}
\caption{
\small
An adaptive continuous attractor neural network (A-CANN) model. 
\textbf{a}. An illustration of the structure of an one-dimensional A-CANN. Neurons are uniformly distributed in the feature space $x\in(-\infty,+\infty)$ to encode a continuous variable (e.g. orientation or direction). Neurons are connected with each
other recurrently with weights $J(x,x')$, and they receive the adaptation current $V(x,t)$ and the external input $I^{ext}(x,t)$.
\textbf{b}. An illustration of the dynamics of a single neuron in the A-CANN. 
\textbf{c}. An illustration of the first two motion modes of a CANN. An arbitrary network state (red curve) can be approximated as the superposition of a steady bump state and the distortions in the bump shape including the height ($v_0$), position($v_1$), and so on.
}
\label{fig1}
\end{figure}

\subsection{An A-CANN model}

We begin with a one-dimensional A-CANN, where neurons are distributed uniformly in a feature space denoted by $x$, with $x$ ranging over the entire real line $(-\infty, \infty)$ (see Fig.~\ref{fig1}a), and neurons are interconnected recurrently. Let $U(x,t)$ be the synaptic input that a neuron at $x$ receives at time $t$, and $r(x,t)$ the corresponding firing rate. The network's dynamics is described as follows:

\begin{eqnarray}	
\tau \frac{dU(x,t)}{dt} & = & -U(x,t)+\rho \int_{-\infty}^{\infty} J(x,x')r(x',t)\, dx'-V(x,t)+I^{ext}(x,t), 
\label{dynamics-U} 
\\
r(x,t) &= &\frac{U(x,t)^2}{1+k\rho \int_{-\infty}^{\infty} U^2(x',t), dx'},
\label{dynamics-r}
\end{eqnarray}
where $\tau$ is the time constant of $U(x,t)$ and $\rho$ the neuron density. The connection strength between neurons at $x$ and $x'$ is denoted as $J(x,x')$, which takes the Gaussian form $J(x,x')=J_0/(\sqrt{2\pi}a)\exp[-(x-x')^2/(2a^2)]$, with $J_0$ controlling the maximum connection strength and $a$ the range of neuronal interactions. The parameter $k$ in Eq.~(\ref{dynamics-r}) controls the amplitude of divisive normalization, which reflects the contribution of inhibitory neurons not explicitly represented in the model. In real biological systems, divisive normalization can be realized via shunting inhibition of inhibitory neurons~\cite{mitchell2003shunting}. It is noteworthy that the neuronal connection profile is translation-invariant in the feature space, i.e., $J(x,x')$ is a function of difference $(x-x')$. This is the key structure of a CANN. $I^{ext}(x,t)$ denotes the external input to the network. 

What distinguishes an A-CANN from a conventional CANN is the term $-V(x,t)$ on the right hand-side of Eq.~(\ref{dynamics-U}). It represents the adaptation current and can be bio-physically traced back to the effect of spike frequency adaptation (SFA) in neuronal dynamics. Although we adopts SFA as the adaptation mechanism in this work, others such as short-term synaptic depression (STD) can also serve as the mechanism.
The dynamics of the adaptation is written as 
(see the illustration in Fig.~\ref{fig1}b),   
\begin{equation}
\tau_v \frac{dV(x,t)}{dt} = -V(x,t)+mU(x,t), \label{dynamics-V}
\end{equation} 	
where $\tau_v$ is the time constant of $V(x,t)$ and $\tau_v\gg \tau$ holds, implying that the adaptation effect caused by neuronal activity is delayed. The parameter $m$ controls the strength of SFA, i.e., the larger the value of $m$ is, the stronger the adaptation effect.


\subsection{The simplified dynamics of the A-CANN}

We introduce how to use a projection method to simplify the dynamics of the A-CANN, which enables us to further carry out theoretical analysis.
Previous studies have shown that because of the localized and translation-invariant recurrent connections, the state of the CANN can be largely approximated as of the Gaussian shape, known as the bump state~\cite{fung2010moving,fung2012dynamical} (see illustration in Fig.~\ref{fig1}c). The bumps are expressed as
\begin{eqnarray}	
	U_0(x,t) & = & A_u \exp \left\{-\frac{\left[x-z(t)\right]^2}{4a^2}\right\}, 
	\label{Ubar}\\
	V_0(x,t) & = & A_v \exp \left\{-\frac{\left[x-z(t)+s_v(t)\right]^2}{4a^2}\right\}, 
	\label{Vbar}\\
	r_0(x,t) & = & A_r \exp \left\{-\frac{\left[x-z(t)\right]^2}{2a^2}\right\},
	\label{rbar}
\end{eqnarray}
where $z(t)$ denotes the bump position, and $A_u$, $A_v$, and $A_r$ denote the heights of bumps $U_0(x,t), V_0(x,t)$, and $r_0(x,t)$, respectively. $s_v(t)$ represents the discrepancy between the neural activity bump $U(x,t)$ and the adaptation current $V(x,t)$. 

The efficacy of the seemingly over-idealistic Gaussian-shaped state comes from that even the bump state is disturbed by noises, the attractor dynamics of the network will rapidly clean out those distortions perpendicular to the attractor space, thereby maintaining the Gaussian-shaped neural activities~\cite{wu2008dynamics,fung2010moving}. Therefore, in below analyses, we assume that Eqs.~(\ref{Ubar}-\ref{rbar}) hold when the external input $I^{ext}(x,t)$ or the adaptation current $V(x,t)$ is small.

Substituting Eqs.~(\ref{Ubar}-\ref{rbar})
into Eqs.~(\ref{dynamics-U}-\ref{dynamics-V}), we have (see Appendix A), 
\begin{eqnarray}
   A_r && = \frac{A_u^2}{1+k\rho\sqrt{2\pi}aA_u^2},
  \label{dynamics_r_eqn}\\
    \tau\left[A_u \frac{x-z}{2a^2}\frac{dz}{dt}+\frac{dA_u}{dt}\right]\mathcal{N}(z,2a) 
    && =  (-A_u+\frac{\rho J_0}{\sqrt{2}}A_r)\mathcal{N}(z,2a) \nonumber \\ 
    && -A_v\mathcal{N}(z-s_v,2a) \nonumber \\ 
    && +I^{ext}(x,t), \label{dynamics_u(t)_eqn} \\
    \tau_v\left[A_v \frac{x-z+s_v}{2a^2}\frac{d(z-s_v)}{dt}+\frac{dA_v}{dt}\right] \mathcal{N}(z-s_v,2a)
    && =  -A_v\mathcal{N}(z-s_v,2a) \nonumber \\
    && +mA_u\mathcal{N}(z,2a), \label{dynamics_v(t)_eqn}
\end{eqnarray}
where $\mathcal{N}(\mu,\sigma)$ denotes the Gaussian function, 
$\mathcal{N}(z,\sigma)=\exp\left[-\left(x-z\right)^2/2\sigma^2 \right]$. At first glance, Eqs.~(\ref{dynamics_u(t)_eqn}-\ref{dynamics_v(t)_eqn}) may appear to be intractable due to the high dimensionality ($2N$, with $N$ the number of neurons). However, a key property of CANNs is that their dynamics are dominated by a few motion modes, which correspond to the distortions of the bump in height, position, width, etc. (see Fig.\ref{fig1}d)~\cite{fung2010moving}. Therefore, by projecting the network dynamics onto these dominant motion modes, we can significantly simplify the network dynamics~\cite{fung2010moving}. Typically, projecting onto the first two motion modes is sufficient to capture the main features of the network dynamics.

For the bump $U(x,t)$, the first two motion modes are,
\begin{eqnarray}	
		  u_0(x,t) & = & \exp \left\{-\frac{[x-z(t)]^2}{4a^2}\right\},
		  \label{u0} \\
		  u_1(x,t) & = & \left[x-z(t)\right]\exp \left\{-\frac{[x-z(t)]^2}{4a^2}\right\}.
		  \label{u1}
\end{eqnarray}

For the bump $V(x,t)$, the first two motion modes are,
\begin{eqnarray}	
	v_0(x,t) & = & \exp \left\{-\frac{[x-z(t)+d(t)]^2}{4a^2}\right\}, 
	\label{v0}\\
	v_1(x,t) & = & \left[x-z(t)+d(t)\right]\exp \left\{-\frac{[x-z(t)+d(t)]^2}{4a^2}\right\}.
	\label{v1}
\end{eqnarray} 

Projecting Eqs.~(\ref{dynamics_u(t)_eqn}-\ref{dynamics_v(t)_eqn}) onto the first two dominant motion modes, we obtain (see Appendix B),
  \begin{eqnarray}
    \tau \frac{dA_u}{dt} & = &  (-A_u+\frac{\rho J_0}{\sqrt{2}}A_r) - A_v\exp \left(-\frac{s_v(t)^2}{8a^2}\right) +\int_{x} I^{ext}(x,t)u_0(x)dx, 
    \label{project_U0_}\\
    \tau A_u \frac{dz}{dt} & = & s_v(t) A_v\exp\left(-\frac{s_v(t)^2}{8a^2}\right)+\int_{x} I^{ext}(x,t)u_1(x)dx,
    \label{project_U1_} \\
     \tau_v \frac{dA_v}{dt} &=& -A_v +mA_u \exp\left(-\frac{s_v(t)^2}{8a^2}\right),
    \label{project_V0_} \\
        \tau_v A_v \frac{d s_v}{dt} & = &   s_v(t) \exp\left(-\frac{s_v(t)^2}{8a^2}\right)\left(\frac{\tau_v}{\tau A_u}A_v^2 - mA_u \right)  + \frac{\tau_v A_v}{\tau A_u}\int_{x} I^{ext}(x,t)u_1(x)dx. \nonumber \\
    \label{project_V1_} 
  \end{eqnarray}
The dimension of the A-CANN dynamics is now reduced to four. In the following sections, we will use these simplified  equations to study the dynamics of A-CANN.

\section{Spontaneous Dynamics of the A-CANN}

In this section, we study the spontaneous dynamics of the A-CANN. By spontaneous dynamics, it refers to the dynamical behaviors of the A-CANN without the drive of external input.


The spontaneous/intrinsic dynamics of neural circuits has gained increasing attention in the field due to its important contributions to information processing in the brain. For instance, the spontaneous activity patterns observed in the developing visual system have been found to play a crucial role in the refinement of neural connections and the establishment of functional circuits~\cite{blankenship2010mechanisms, chiu2001spontaneous}. Similarly, ongoing oscillations in the brain have been linked to important cognitive processes such as attention, perception, and memory~\cite{ward2003synchronous}. 
An appealing substrate for these oscillations is the spontaneous hippocampal replay referring to the reactivation and replay of neural activity patterns in the hippocampus. It occurs during the rest or sleep period (offline) and involves sequential reactivation of specific neurons or groups of neurons that were active during previous awake experiences. It is suggested that the hippocampal replay plays a crucial role in memory consolidation and spatial navigation, as it supports the strengthening and integration of newly acquired experiences~\cite{dupret2010reorganization, carr2011hippocampal}. Through replay, most heavy computations of constructing representation space take place offline, thus the computational burden is reduced in online behaviors.

Moreover, the study of spontaneous activity is crucial in understanding neurological and psychiatric disorders, as abnormal patterns of intrinsic activity have been observed in conditions like epilepsy, schizophrenia, and autism spectrum disorders~\cite{northoff2016abnormalities, chang2019spontaneous}. Analyzing these spontaneous dynamics can provide valuable insights into the underlying neural mechanisms of such disorders and aid in developing potential therapeutic strategies.

\subsection{Static bump state}
We first study the condition when the A-CANN holds static bumps as its stationary states. When no external input exists, the network dynamics given by Eqs.~(\ref{project_U0_}-\ref{project_V1_}) are further simplified to be,
\begin{eqnarray}
    \frac{dA_u}{dt} & = & \frac{1}{\tau } \left(-A_u+\frac{\rho J_0}{\sqrt{2}}A_r - A_v\right), 
    \label{project_U0_static}\\
    \frac{dz}{dt} & = &\frac{s_v(t) A_v}{\tau A_u },
    \label{project_U1_static} \\
     \frac{dA_v}{dt} &=& \frac{1}{\tau_v} \left(-A_v +mA_u \right),
    \label{project_V0_static} \\
    \frac{ds_v(t)}{dt} & = &  s_v(t) \left(\frac{A_v}{\tau A_u} - \frac{mA_u}{\tau_v A_v } \right).
    \label{project_V1_static} 
\end{eqnarray}
In the above, we have applied the condition $s_v(t) << a$ which gives  $\exp\left[-s_v(t)^2/8a^2\right]\approx 1$. This approximation holds well in practice.

In the static bump state, the bump center remains stationary, i.e., $dz/dt = 0$, $s_v(t) = 0$, and bump heights are constant, i.e., $dA_u/dt = dA_v/dt = dA_r/dt = 0$. Substituting these constraints into Eqs.(\ref{project_U0_static}-\ref{project_V1_static}), we have,
\begin{eqnarray}
    (-A_u+\frac{\rho J_0}{\sqrt{2}}A_r) - A_v & = & 0 , \\
    -A_v +mA_u &=& 0. 
\end{eqnarray}
Combining with Eq.~(\ref{dynamics_r_eqn}), we obtain,
\begin{eqnarray}
    A_u&=&\frac{\rho J_0 +\sqrt{\rho^2J_0^2-8\sqrt{2\pi}(1+m)^2k\rho a}}{4\sqrt{\pi}(1+m)k\rho a}.
    \label{Au_static_bump}\\
   A_v&=&mA_u,\label{Av_static_bump}\\
   A_r& = & \frac{\sqrt{2}(1+m)}{\rho J_0}A_u.
    \label{Ar_static_bump}
\end{eqnarray}
Thus, the static bump state is solved. To
verify the theoretical result, we conducted simulations of the original network dynamics. Eq.~(\ref{Au_static_bump}) predicts that
increasing the global inhibition $k$ decreases the the height of the neural activity bump $A_u$. Fig.~\ref{fig2}a shows that our theoretical analysis agrees very well with the simulation.
 
We proceed to analyze the stability of the static bump state by examining how the global inhibition strength $k$ and adaptation strength $m$ impact the bump state. A widely used and powerful tool in computational neuroscience for studying the dynamics of complex systems is Jacobian matrix, which provides valuable information about the local behavior of a system at certain point in phase space. By analyzing the eigenvalues of Jacobian matrix, we can determine the stability of the state and classify it as a fixed point, a limit cycle, or a chaotic attractor. The signs of the eigenvalues reveal the motion tendency of the system under small perturbations, indicating whether they will cause the system to return to its steady-state or diverge from it. Specifically, we calculate the eigenvalues of the matrix derived from Eqs.(\ref{project_U0_static} - \ref{project_V1_static}). The Jacobian matrix is calculated to be:
\begin{equation}      
\textbf{M}=\left(                
  \begin{array}{cccc}  
    \frac{1}{\tau} \left(-1+\frac{\sqrt{2}\rho J_0 A_u }{(1+\sqrt{2\pi}k \rho a A_u^2)^2}\right) & -\frac{1}{\tau} & 0 & 0\\ 
    \frac{m}{\tau_v} & -\frac{1}{\tau_v} & 0 & 0 \\  
    -\frac{s_v(t)A_v}{\tau A_u^2} & \frac{s_v(t)}{\tau A_u} & 0 & \frac{A_v}{\tau A_u} \\
    -\frac{s_v(t)A_v}{\tau A_u^2} & \frac{s_v(t)}{\tau A_u} & 0 & \left(\frac{A_v}{\tau A_u} - \frac{mA_u}{\tau_v A_v } \right) \\
  \end{array}
\right).  
\end{equation}
Considering the condition $s_v(t) \approx 0$ near the steady state, we observe that the matrix becomes block-diagonal, where the dynamics of $A_u(t)$ and $A_v(t)$ are completely independent of the dynamics of $z(t)$ and $s_v(t)$. Therefore, we can simplify the calculation by separating the 4-rank matrix into two smaller 2-rank matrices. 

First, we focus on the 2-rank Jacobian matrix of the dynamics of $A_u(t)$ and $A_v(t)$, which yields:
\begin{equation}      
\textbf{M}_1=\left(                
  \begin{array}{cc}  
    \frac{1}{\tau} \left(-1+\frac{\sqrt{2}\rho J_0 A_u }{(1+\sqrt{2\pi}k \rho a A_u^2)^2}\right) & -\frac{1}{\tau} \\ 
    \frac{m}{\tau_v} & -\frac{1}{\tau_v} \\  
  \end{array}
\right). 
\end{equation}
Denote the eigenvalues of the Jacobian matrix as $\lambda_1$ and $\lambda_2$. For the solution to be stable, both eigenvalues need to be negative, which require that,
\begin{eqnarray}
    \lambda_1+\lambda_2&=&\frac{1}{2}\left[-1+\frac{\sqrt{2}A_uJ_0\rho} {(1+\sqrt{2\pi}k \rho a A_u^2)^2}-\frac{\tau}{\tau_v}\right]<0,\\
    \lambda_1 \lambda_2&=&\frac{\tau}{\tau_v}\left(m+1-\frac{\sqrt{2}A_uJ_0\rho} {(1+\sqrt{2\pi}k \rho a A_u^2)^2}\right)\geq 0.
\end{eqnarray}
The above inequalities are satisfied when,
\begin{eqnarray}
    0<k<k_{c1}&=&\frac{\rho J_0^2(1+\frac{\tau}{\tau_v})(1+2m-\frac{\tau}{\tau_v})}{8\sqrt{2\pi}a(1+m)^4},\\
    0<k<k_{c2}&=&\frac{\rho J_0^2}{8\sqrt{2\pi}a(1+m)^2}.
    \label{k-boundary}
\end{eqnarray}
It is easy to check that $k_{c2}<k_{c1}$. Thus, the first condition for the network to hold static bumps as its steady state is $0<k<k_{c2}$.

Second, we study the 2-rank Jacobian matrix of the the dynamics of $z(t)$ and $s_v(t)$, which gives,
\begin{equation}      
\textbf{M}_2=\left(                
  \begin{array}{cc}  
    0 & \frac{m}{\tau} \\ 
    0 & \frac{m}{\tau}-\frac{1}{\tau_v}
  \end{array}
\right).  
\end{equation}
It's easy to check that the eigenvalues are $\lambda_1 = 0, \lambda_2 = m-\tau/\tau_v$. Thus, the second condition for the static bump state to be stable is $m<\tau/\tau_v$.

In summary, the parametric conditions for the A-CANN holding a static bump as its stationary state without relying on external input are that
both the inhibition and adaptation need to be sufficiently small, satisfying $0<k<k_{c2}$ and $m<\tau/\tau_v$. These agree with our intuition, as too strong inhibition will suppress neuronal activities and
too large adaptation will destabilize the static bump (see below).

\subsection{Traveling wave state}
\label{Traveling_wave}
\begin{figure}[H]
\hfill
\begin{center}
\includegraphics[width=14 cm]{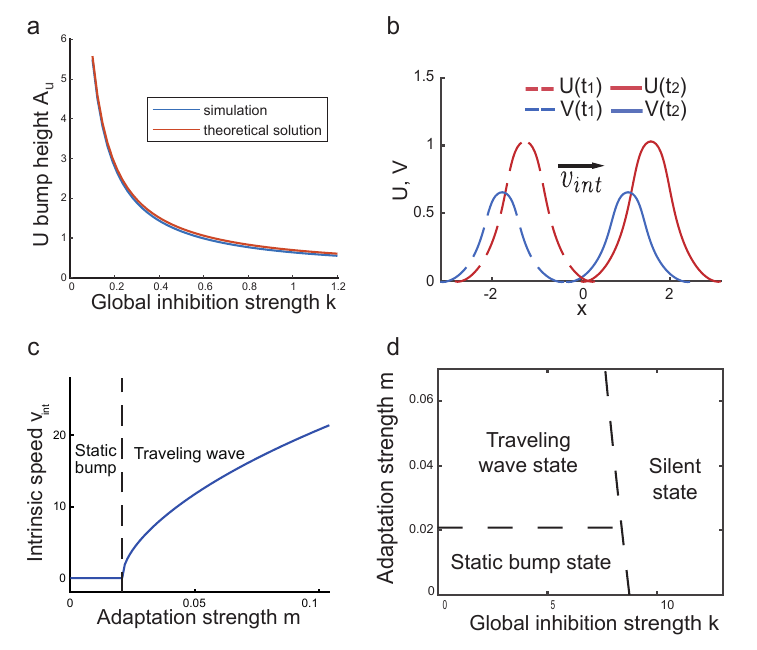}
\end{center}
\small
\caption{The spontaneous dynamics of the A-CANN. 
\textbf{a}. The bump height $A_u$ vs. the global inhibition strength $k$.
\textbf{b}. An illustration of the travelling wave state of the A-CANN. The bumps $U$ (red) and $V$ (blue) moves spontaneously with the speed $v_{int}$, and the discrepancy between their centers is $s$.
\textbf{c}. As the adaptation strength $m$ increases across a threshold $m_0=\tau/\tau_v$, the steady state of the A-CANN changes from static bump ($v_{int}=0$) to travelling wave ($v_{int}>0$). 
\textbf{d}. The phase diagram of spontaneous states of the A-CANN. Parameters are: $a = 0.4$, $J_0 = 1$, $k = 0.76$,
$\tau= 3ms$,$\tau_v= 152ms$, $\rho = 128/(2\pi)$, $m = 0.3$, $\alpha = 0.2$, $v_{ext}=0.5/s$.
}\label{fig2}
\end{figure}

The above analysis reveals that for the A-CANN holding static bump states, it requires the adaptation strength $m<\tau/\tau_v$.
If $m>\tau/\tau_v$, the strong adaptation will destabilize the bump,
causing the bump to move spontaneously in the feature space (illustrated in Fig.\ref{fig2}b). This is called the traveling wave state~\cite{bressloff2011spatiotemporal, mi2014spike, coombes2014neural, wu2016continuous}.
In this section, we study the condition for the A-CANN holding the traveling wave state. 

Let us denote the speed of travelling wave to be $v_{int}$,
which is also called the intrinsic speed of the A-CANN, as its value
only depends on the network parameters. Without loss of generality, the bump center can be expressed as $z(t) = v_{int}t$.
At the travelling wave state, the bump heights remain invariant, i.e., $dA_u/dt = dA_v/dt = 0$. The network dynamics Eqs.~(\ref{project_U0_}-\ref{project_V1_}) are simplified to be,
  \begin{eqnarray}
    A_u & = &  \frac{\rho J_0}{\sqrt{2}}A_r - A_v\exp \left(-\frac{s_v(t)^2}{8a^2}\right) ,
    \label{project_U0}\\
    \tau A_u v_{int} & = & s_v(t) A_v\exp\left(-\frac{s_v(t)^2}{8a^2}\right),
    \label{project_U1} \\
     A_v &=& mA_u \exp\left(-\frac{s_v(t)^2}{8a^2}\right),
    \label{project_V0} \\
    \tau_v A_v \frac{ds_v(t)}{dt} & = &  s_v(t) \exp\left(-\frac{s_v(t)^2}{8a^2}\right)\left(\frac{\tau_v}{\tau A_u}A_v^2 - mA_u \right).
    \label{project_V1} 
  \end{eqnarray}

Combining the above equations with Eq.~(\ref{dynamics_r_eqn}), we can analytically solve the travelling wave state, which are given by (Appendix C):
  \begin{eqnarray}
  A_u & =& \frac{\rho J_0+\sqrt{\rho^2J_0^2-8\sqrt{2\pi}k\rho a(1+\sqrt{\frac{m\tau}{\tau_v}})^2}}
  {4\sqrt{\pi}k\rho a(1+\sqrt{\frac{m\tau}{\tau_v}})}, \label{A_u_theo}\\
  A_v & =& \frac{\rho J_0+\sqrt{\rho^2J_0^2-8\sqrt{2\pi}k\rho a(1+\sqrt{\frac{m\tau}{\tau_v}})^2}}{2\sqrt{2 \pi }k \rho ^2 a J_0}, \\
  A_r & =& \sqrt{\frac{m\tau}{\tau_v}}
  \exp\left[\frac{1-\sqrt{\frac{\tau}{m\tau_v}}}{2}\right]
  \frac{\rho J_0+\sqrt{\rho^2J_0^2-8\sqrt{2\pi}k\rho a(1+\sqrt{\frac{m\tau}{\tau_v}})^2}}{4\sqrt{\pi}k\rho a(1+\sqrt{\frac{m\tau}{\tau_v}})} , \\
  s_v &=& 2a\sqrt{1-\sqrt{\frac{\tau}{m\tau_v}}}, \\
  v_{int} &=& \frac{2a}{\tau_v}\sqrt{\frac{m\tau_v}{\tau}-\sqrt{\frac{m\tau_v}{\tau}}}.\label{v_int}
  \end{eqnarray}
Thus, the travelling wave speed is given by $v_{int}=a/\tau_v\sqrt{m\tau_v/\tau-\sqrt{m\tau_v/\tau}}$, and the lag of the adaptation current to the bump center $s_v$ remain a constant.  Fig.~\ref{fig2}c confirms that theoretical prediction of the intrinsic speed $v_{int}$ agrees well with the simulation result. From Eq.~(\ref{v_int}), we also observe that for the traveling wave speed $v_{int}$ to take a real value, it requires, 
\begin{equation}
m>\frac{\tau}{\tau_v},
\label{m_boundary}
\end{equation}
which is consistent with the stability analysis in the previous section. 
Combining Eqs.~(\ref{k-boundary},\ref{m_boundary}), we can draw the phase
diagram of the spontaneous states of the A-CANN, which is depicted in
Fig.\ref{fig2}d.

\subsubsection{Biological implications of travelling wave}
Spontaneous propagation of neural activity has been widely observed in both {\it in vitro} and {\it in vivo} experiments conducted on the cortex ~\cite{luhmann2016spontaneous, mitra2016networks} and sub-cortical areas like hippocampus ~\cite{herreras1994role, both2008propagation}. 
Here, our model suggests a mechanism to generate these spontaneous activities as a consequence of neural adaptation.  
Consider a bump emerges randomly 
at a location in the attractor space due to noises. 
Because of the adaptation, those neurons around the bump position (i.e., those most active neurons) receive the strongest suppression, which destabilizes the bump at the current location. Aided by recurrent interactions between neurons, the bump moves away to the neighborhood, where the adaptation starts to destabilize the bump at the new location again. Consequently, the bump keeps moving in the attractor space, and the network exhibits the travelling wave behavior. 

The traveling wave behavior may play important roles in neural information processing. Intuitively, it enables a neural system
to progressively visit all stationary states, and hence endows the neural system with the capacity of actively retrieving all stored memories. This could support cognitive functions where memory search and memory consolidation are involved~\cite{carr2011hippocampal}. Tracing computational roles of traveling wave could be a fascinating avenue for future research.

\unskip



\section{Tracking Dynamics of the A-CANN}
In the above, we have studied the spontaneous dynamics of the A-CANN in the absence of external inputs. In this section, we further study the tracking dynamics of the A-CANN when an external moving input is presented. When no adaptation exists, the continuous manifold of attractor states enables a conventional CANN to track a moving input smoothly with a constant delay~\cite{fung2010moving,wu2008dynamics,wu2005computing}.
When adaptation is included, it induces rich tracking behaviors of the network~\cite{mi2014spike, wong2010attractor}.

Without loss of generality, we consider the external moving input taking the below form,
\begin{equation}	
	I^{ext}(x,t)=\alpha \exp \left[-\frac{(x-v_{ext}t)^2}{4a^2}\right],
	\label{I^{ext}}
  \end{equation} 
where $v_{ext}$ is the speed of the external input and $\alpha$ controls the strength of the input. 

For the convenience of analysis, we assume that the bump heights remain constants during the tracking process, i.e., $dA_u/dt=0, dA_v/dt=0, dA_r/dt=0$ (this assumption is not always feasible as in the case of oscillatory tracking, nevertheless, it gives good approximation results). Substituting Eq.~(\ref{I^{ext}}) and constant amplitude assumption into the network dynamics Eqs.~(\ref{project_U0_}-\ref{project_V1_}),  we obtain the simplified tracking dynamics, which are:
\begin{eqnarray}
	A_u& = & \frac{\rho J_0}{\sqrt{2}}A_r+\alpha \exp\left(-\frac{s_u(t)^2}{8a^2}\right)-A_v \exp\left(-\frac{s_v(t)^2}{8a^2}\right), 
	\label{project_U0_tracking}\\
        A_v&=&mA_u \exp\left(-\frac{s_v(t)^2}{8a^2}\right),
	\label{project_U1_tracking}\\
  \frac{ds_u(t)}{dt}& = & \frac{A_v}{\tau A_u}s_v(t) \exp\left(-\frac{s_v(t)^2}{8a^2}\right)-\frac{\alpha }{\tau A_u}s_u(t)\exp\left(-\frac{s_u(t)^2}{8a^2}\right)-v_{ext},
	\label{project_V0_tracking}\\
		\frac{ds_v(t)}{dt}&=&\left(\frac{A_v}{\tau A_u}-\frac{mA_u}{\tau_v A_v}\right)s_v(t) \exp\left(-\frac{s_v(t)^2}{8a^2}\right)-\frac{\alpha }{\tau A_u}s_u(t)\exp\left(-\frac{s_u(t)^2}{8a^2}\right), \nonumber \\
	\label{project_V1_tracking}
\end{eqnarray}
where $s_u(t)$ denotes the discrepancy between the neural activity and the external input bumps, i.e., $s_u(t) = z(t) - v_{ext}t$. 


In the blow study, we focus on the case that the speed of the external input complies with the condition $v_{ext} \ll a/\tau_v$.
This is the biologically relevant regime in practice. For instance, if to model head rotation with biologically plausible parameters such as $\tau_v = 100 ms$ and $a = 50^\circ$, $v_{ext} = 0.1a/\tau_v$ corresponds to a speed of $500^\circ/s$, which is quite fast for head rotation of rodents. Furthermore, we take the approximations $\exp(-s_v^2/8a^2) \approx 1$ and $\exp(-s_u^2/8a^2) \approx 1$, as they always hold in practice. 
From Eqs.~(\ref{project_U0_tracking}-\ref{project_U1_tracking}), we have 
\begin{equation}	
  (m+1)A_u-\frac{\rho J_0}{\sqrt{2}}\frac{A_u^2}{1+\sqrt{2\pi}ak\rho A_u^2}-\alpha=0.
  \label{cubic_equation}
\end{equation} 
Eq.~(\ref{cubic_equation}) can be rearranged into a general cubic equation of $A_u$, which is written as,
\begin{equation}
    a_3 A_u^3 + a_2 A_u^2 + a_1 A_u + a_0 = 0,
    \label{cubic_equation_Au}
\end{equation}    
where $a_3 = \sqrt{2\pi}(m+1)ak\rho$,
$a_2 = -\sqrt{2\pi}ak\rho\alpha-\rho J_0/\sqrt{2}$,
$a_1 = m+1$, and $a_0 = -\alpha$.
It is easy to check that Eq.~(\ref{cubic_equation_Au}) only has one real solution, which is written as,
\begin{equation}    
    A_u = \left[-\frac{q}{2}+\sqrt{(\frac{q}{2})^2+(\frac{p}{3})^3}\right]^{1/3}+\left[-\frac{q}{2}-\sqrt{(\frac{q}{2})^2+(\frac{p}{3})^3}\right]^{1/3},
    \label{cubic_solution_Au}
\end{equation}
where  $q = (3a_3a_1-a_2^2)/(3a_3^2)$ and 
$p = (27a_3^2a_0-9a_3a_2a_1+2a_2^3)/(27a_3^3)$.
The form of $A_u$ given by Eq.~(\ref{cubic_solution_Au}) is quite complicated. By numerical simulation, we find that the condition $\sqrt{2\pi}ak\rho A_u^2 \gg 1$ always holds. This gives
$A_u^2/(1+\sqrt{2\pi}ak\rho A_u^2) \approx 1/\sqrt{2\pi}ak\rho$. Thus, the expression of $A_u$ can be simplified to be,
\begin{equation}	
    A_u=\frac{J_0+2\sqrt{\pi}ak\alpha}{2\sqrt{\pi}ak(1+m)}.
    \label{Expression for A_u}
\end{equation} 		
From Eq.~(\ref{project_U1_tracking}), we have $A_v\approx m A_u$. 
In the below, we continue to solve the dynamics of $s_u(t)$ and $s_v(t)$, which vary with different parameter conditions. 

\begin{figure}[H]
\hfill
\begin{center}
\includegraphics[width=15 cm]{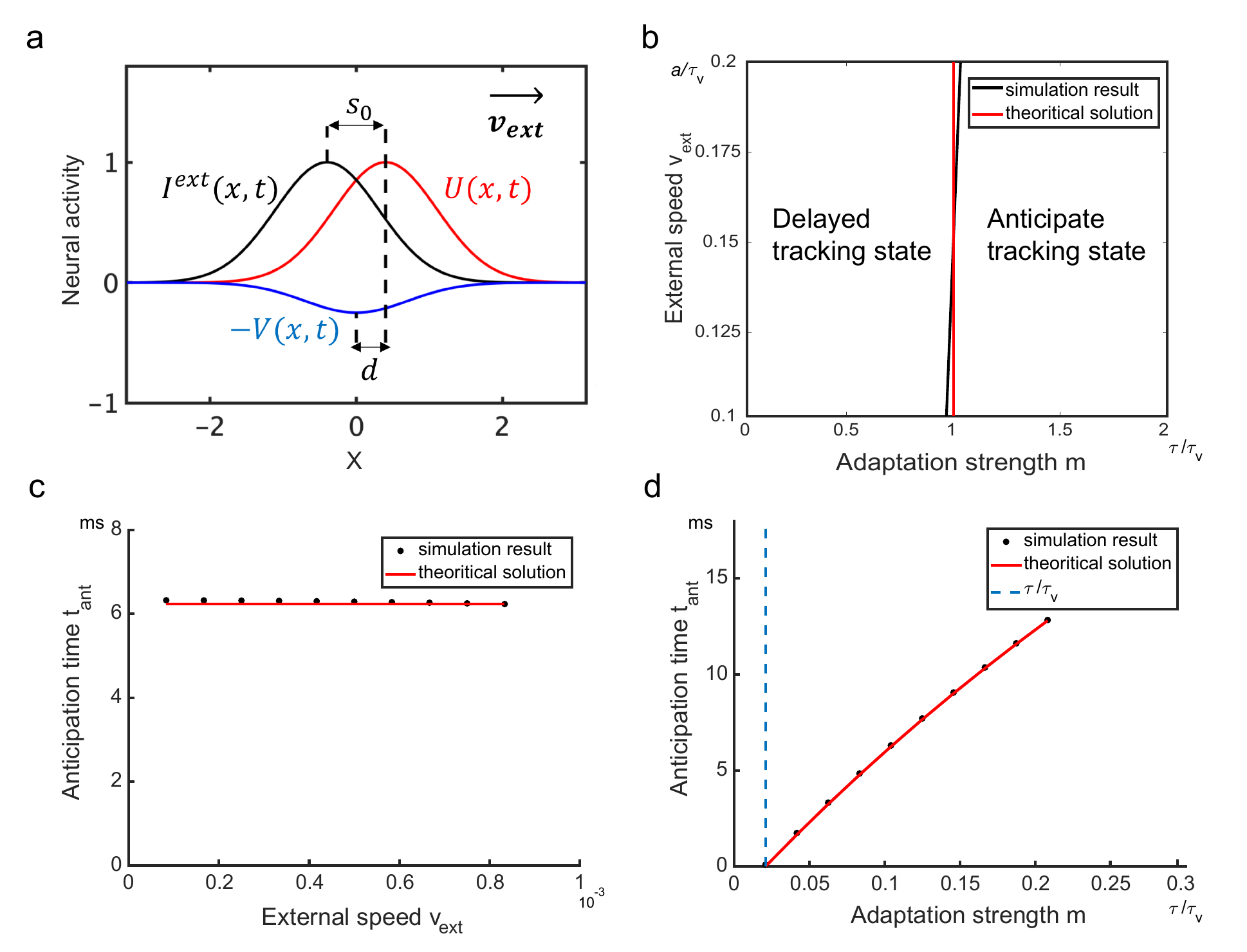}
\end{center}
\caption{
\small
Smooth tracking of the A-CANN. 
\textbf{a}. An illustration of the smooth tracking state of A-CANN. The neural activity bump $U(x,t)$ moves at the speed $v_{ext}$ of the external input $I^{ext}(x,t)$ with a separation $s_u$. The adaptation current $V(x,t)$ is delayed by $s_v$ with respect to the bump $U(x,t)$.
\textbf{b}. Depending on the relative sizes of $v_{ext}$ and $v_{int}$ (scaled by $\tau/\tau_v$), the bump can either lead the external input ($s_u>0$, anticipate tracking state) or lag the external input ($s_u<0$, delayed tracking state).
\textbf{c}. For small speed values, the A-CANN achieves a roughly constant anticipation time $t_{ant}$ with respect to the external input.
\textbf{d}. The anticipation time $t_{ant}$ increases roughly linearly with adaptation strength $m$.
Parameters are: $a = 0.4$, $J_0 = 1$, $k = 5$, $\tau= 1ms$, $\tau_v= 48ms$, $\rho = 512/(2\pi)$, $\alpha = 0.19$, $m = 0.1$ in (\textbf{b}), $v_{ext}=0.5/s$ in (\textbf{c}).
}\label{fig3}
\end{figure}   

\subsection{Smooth tracking state}
We begin by studying the smooth tracking state, in which the neural bump tracks the external input in a phase-locking manner, in terms of that
the discrepancy between the neural activity bump and the external input bump $s_u(t)$, and the discrepancy between the neural activity bump and the adaptation current bump $s_v(t)$ are constants, i.e., 
$ds_u(t)/dt=0$ and $ds_v(t)/dt=0$. This occurs when the external input strength is sufficiently strong or the adaptation strength is sufficiently weak.   

Substituting $ds_u(t)/dt=0, ds_v(t)/dt=0$ into Eqs.~(\ref{project_V0_tracking}-\ref{project_V1_tracking}), we have,
\begin{eqnarray}
		s_u\exp\left(-\frac{s_u^2}{8a^2}\right)&=&\frac{\tau A_u v_{ext}}{\alpha}\left[\frac{\tau_v}{m\tau}\left(\frac{A_v}{A_u}\right)^2-1\right],
		\label{su_dynamics}\\
		s_v\exp\left(-\frac{s_v^2}{8a^2}\right)&=&\frac{\tau_v A_v v_{ext}}{mA_u}.
		\label{sv_dynamics}
\end{eqnarray}
Combining Eqs.~(\ref{su_dynamics}-\ref{sv_dynamics}) with Eqs.~(\ref{project_U0_tracking}-\ref{project_U1_tracking}) , we solve the smooth tracking state, which gives, 
\begin{eqnarray}
	s_v&=&\tau_v v_{ext},\label{Expression for s_v}\\
	s_u&=&\frac{\tau_v v_{ext}\left[m\exp \left(-\frac{\tau_v^2 v_{ext}^2}{4a^2} \right) -\frac{\tau}{\tau_v} \right]}{1-\frac{\rho J_0 A_r}{\sqrt{2}A_u}-m\left(\exp \left(-\frac{\tau_v^2 v_{ext}^2}{4a^2} \right)\right)^2}.
	\label{Expression for s_u}
\end{eqnarray}
It can be checked from Eq.~(\ref{Expression for s_u}) that 
when $v_{int}>v_{ext}$, $s_u>0$, that is, the neural activity bump leads the external input.

Furthermore, when the conditions $v_{ext}\ll a/\tau_v$ and $s_u\ll 2\sqrt{2}a$ are satisfied, the value of $s_u$ can be approximated to be 
\begin{equation}	
	s_u\approx\frac{A_u v_{ext}\tau_v}{\alpha}(m-\frac{\tau}{\tau_v}),
 \label{eq-su}
  \end{equation}   
which indicates that $s_u$ increases linearly with $v_{ext}$. In such a case, the anticipation time is a constant independent of the input speed, which is calculated to be 
\begin{equation}	
	t_{ant}=\frac{s}{v_{ext}}\approx\frac{A_u\tau_v}{\alpha}(m-\frac{\tau}{\tau_v}). \label{Expression for t_ant}
\end{equation} 
These theoretical results are verified by simulations (Fig.\ref{fig3}b-c). 

In summary, we have two observations: 1) the neural bump of
the A-CANN can track the external moving input in an anticipative manner
when $v_{int}>v_{ext}$; 2) the anticipation time can be approximated as a constant for a wide range of input speeds. 

\subsubsection{Stability analysis of the smooth tracking state}
We proceed to study the stability of the smooth tracking state.
The Jacobian matrix of the simplified tracking dynamics in Eqs.~(\ref{project_V0_tracking}-\ref{project_V1_tracking}) is calculated to be:
\begin{equation}
\textbf{M}=\left( 
  \begin{array}{cc}
    -\frac{\alpha}{\tau A_u} F(s_u) & \frac{A_v}{\tau A_u} F(s_v) \\
   -\frac{\alpha}{\tau A_u} F(s_u) & \left(\frac{A_v}{\tau A_u}-\frac{mA_u}{\tau_v A_v}\right) F(s_v)
  \end{array}
\right),
\end{equation}
where the function $F(x)=(1-x^2/4a^2)\exp(-x^2/8a^2)$. The matrix can be further simplified as,
\begin{equation}
\textbf{M}=\frac{A_v}{\tau A_u}F(s_u)\left( 
  \begin{array}{cc}
    -\phi & 1 \\
   -\phi & (1-\theta)
   \label{Jacobian_Matrix}
  \end{array}
\right),
\end{equation}
in which,
\begin{eqnarray}
	\theta &=&\frac{m\tau}{\tau_v}(\frac{A_u}{A_v})^2,
		\label{theta}\\
	\phi &=&\frac{\alpha}{A_v}\frac{F(s_u)}{F(s_v)}.
		\label{phi}
\end{eqnarray}
According to the Vieta theorem, the eigenvalues of the Jacobian matrix $\bf{M}$, denoted as $\lambda_1$ and $\lambda_2$, satisfy,
\begin{eqnarray}
    \lambda_1+\lambda_2&=& 1-\theta-\phi,\\
    \lambda_1 \lambda_2&=& \phi \theta >0.\label{product}
\end{eqnarray}
For the smooth tracking state to be stable, it requires that both eigenvalues have negative real parts, which implies:
\begin{equation}
    \theta + \phi > 1.
    \label{Stationary_condition}
\end{equation}
We consider that the external input speed $v_{ext}\ll a/\tau_v$ and the discrepancies $s_u\ll a$ and $s_v\ll a$. Under these conditions, we have the approximations $F(s_u) \approx F(s_v) \approx 1$ and $A_v \approx mA_u$ (see Eq.~(\ref{project_U1_tracking})). Substituting Eqs.~(\ref{theta}-\ref{phi}, \ref{project_U1_tracking}) into Eq.~(\ref{Stationary_condition}), we get,
\begin{equation}
    m-\frac{\tau}{\tau_v} < \frac{\alpha}{A_u}. 
    \label{Stationary_condition_approx}
\end{equation}
Based on Eq.~(\ref{Stationary_condition_approx}), we differentiate two different situations on the stability of the smooth tracking state:
1) when the adaptation strength is too weak to produce a traveling wave, i.e., $m<\tau/\tau_v$, the smooth tracking state is stable 
since Eq.~(\ref{Stationary_condition_approx}) always holds (note $\alpha\geq 0$). 2) When the adaptation strength $m>\tau/\tau_v$, the smooth tracking state is stable only when $m<\tau/\tau_v+\alpha/A_u$, i.e., it requires the external input strength $\alpha$ to be sufficiently large.

\subsubsection{Biological implications of anticipative tracking}
A CANN can track an external moving input smoothly, a property that has been used to model spatial navigation in the hippocampus~\cite{samsonovich1997path}. The tracking of a CANN is, however, always lagging behind the true location of the external input, as it takes time for the network responding to the external input change. Here, we show that by including strong enough adaptation in the neural dynamics, the A-CANN is able to track an external moving input anticipatively. The underlying mechanism can be intuitively understood as follows. The adaptation induces intrinsic mobility of the neural bump, in term of that the bump tends to move spontaneously as travelling wave at the speed $v_{int}$ when the adaptation strength is sufficiently strong, while the external moving input wants to drive the bump to moves at the speed $v_{ext}$. These two forces compete with each other, leading to that the bump leads the external input when $v_{int}>v_{ext}$. Furthermore, if the speed of the external input is not too large, the bump leads the external input with a constant leading time.  

Time delays are pervasive and significant in the brain, as the transmission of neural signal over hierarchical layers take time. If these delays are not compensated properly, our perception and reaction will lag behind the real circumstance in the external world significantly, impairing our vision and movement. The anticipative tracking behavior of the A-CANN
may provide a simple yet effective mechanism to compensate these delays.
For instance, the experiment depicted that the internal head-direction representation in the anterior thalamus of a rat anticipates the animal's head-direction by about $20-25$ms, independent of the rotation speed of the head~\cite{blair1995anticipatory}. The head-direction system has been modeled using CANNs~\cite{zhang1996representation}, and the A-CANN can justify the constant anticipation time as observed in the experiment. 

\subsection{Oscillatory tracking state}

\begin{figure}[H]
\hfill
\begin{center}
\includegraphics[width=15 cm]{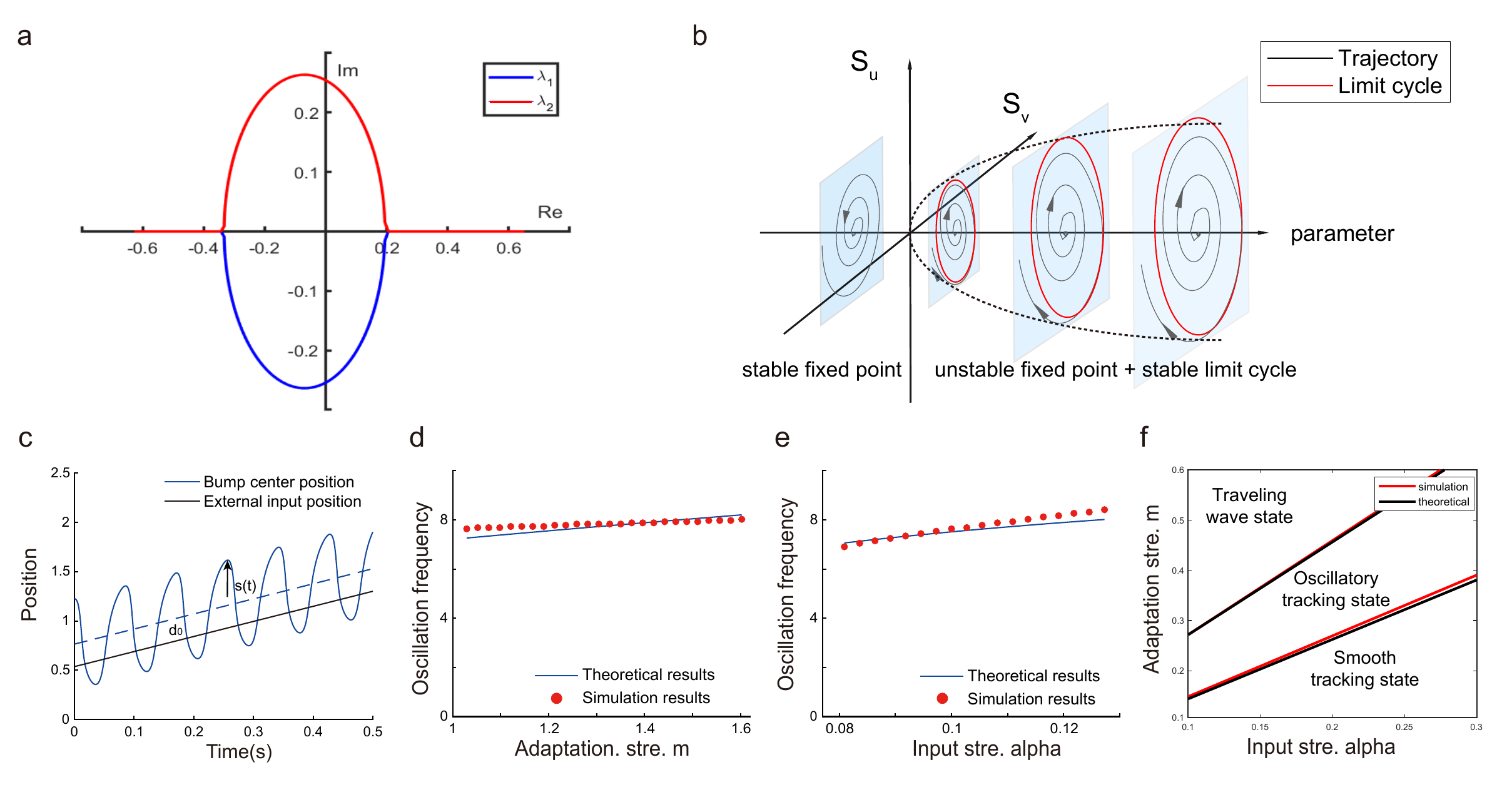}
\end{center}
\small
\caption{Oscillatory tracking of the A-CANN. 
\textbf{a}. Eigenvalues cross the imaginary axis from a non-zero point.
\textbf{b}. Geometric interpretation of the Hopf bifurcation and the emergence of oscillatory tracking. The Hopf bifurcation is generated when the eigenvalues cross the imaginary axis, which leads to oscillation (red line).
\textbf{c}. The oscillatory tracking process of the A-CANN. The bump center (blue dashed line) tracks the position of the external input (black line) in an  oscillatory way with a constant offset $d_0 > 0$. The parameters are chosen, such that $\omega\approx 8 Hz$ is in  the range of theta frequency.
\textbf{d}-\textbf{e}. The oscillation frequency increases approximately linearly over a wide range of $m$ (\textbf{d}) and $\alpha$ (\textbf{e}).
\textbf{f}. Phase diagram of the tracking dynamics of the A-CANN.
Parameters are: $a = 0.4$, $J_0 = 1$, $k = 0.76$,
$\tau= 3ms$,$\tau_v= 152ms$, $\rho = 128/(2\pi)$, $m = 0.3$, $\alpha = 0.2$, $v_{ext}=0.5/s$.
}\label{fig4}
\end{figure}

We further study the tracking dynamics of the A-CANN when the smooth tracking state is unstable. Eq.~(\ref{Stationary_condition_approx}) 
shows that the bifurcation point is at $m_0 =\alpha A_u + \tau/\tau_v$.
As $m$ increases from a small value to be larger than $m_0$, the real parts of eigenvalues $\lambda_1$ and $\lambda_2$ change sign from negative to positive. At the bifurcation point, the phase space formed by $ds_u/dt$ and $ds_v/dt$ changes its topology.
The type of bifurcation depends on the imaginary parts of eigenvalues at the fixed point, which is given by,
\begin{equation}
    \Delta = \sqrt{(1-\theta-\phi)^2-4\phi\theta}.
\end{equation}
At the bifurcation point, $\lambda_1 + \lambda_2 = 1-\theta-\phi = 0$ holds, which leads to $\Delta^2 = -4\phi\theta < 0 $. As a result, the bifurcation point $m_0$ is Hopf bifurcation, in term of that the eigenvalues change their signs through the imaginary axis (see illustration in Fig.\ref{fig4}b)~\cite{luenberger1979introduction}.
Hopf bifurcation incurs the emergence of limit cycle dynamics of a dynamical system (see Fig.~\ref{fig4}a)~\cite{luenberger1979introduction}. This leads to the neural bump of the A-CANN tracks the external input in an oscillatory way, in term of that the neural bump will sweep forth and back around the external input~\cite{coombes2014neural, folias2004breathing, chu2022oscillatory} (Fig.~\ref{fig4}c).

For the convenience of analysis, in the oscillatory tracking state, we assume that the discrepancies $s_u(t)$ and $s_v(t)$ undergo harmonic oscillations, which are expressed as:
\begin{eqnarray}
    s_u(t)=c_u sin(\omega t)+d_u,
    \label{osc_su}\\
    s_v(t)=c_v sin(\omega t + \gamma)+d_v,
    \label{osc_sv}
\end{eqnarray}
where $c_u, c_v$ and $\omega$ represent the amplitudes and frequency of the bump oscillation, $d_u$ and $d_v$ represent the average offsets between the bump oscillation and the external input, $\gamma$ represents the phase offset between the bumps of $U(x,t)$ and $V(x,t)$. 

Substituting Eqs.~(\ref{osc_su}-\ref{osc_sv}) and (\ref{dynamics_r_eqn}) into the tracking dynamics given by Eqs.~(\ref{project_U0_tracking}-\ref{project_V1_tracking}), and further taking the approximations that $\exp(-s_u^2/8a^2) \approx \exp(-s_v^2/8a^2) \approx 1$ and that bump heights $A_u, A_v$ and $A_r$ are constants, we get,
\begin{eqnarray}
	A_u& = & \frac{\rho J_0}{\sqrt{2}}\frac{A_u^2}{1+k\rho\sqrt{2\pi}aA_u^2}+\alpha-A_v, 
	\label{project_U0_osc}\\
        A_v&=&mA_u,
	\label{project_U1_osc}\\
  c_u\omega \cos{\omega t}& = & \frac{A_v}{\tau A_u}(c_v\sin{(\omega t + \gamma)}+d_v)-\frac{\alpha }{\tau A_u}(c_u\sin{\omega t}+d_u)-v_{ext},
	\label{project_V0_osc}\\
		c_v\omega \cos{(\omega t+\gamma)}&=&-\frac{mA_u}{\tau_v A_v}(c_v\sin{(\omega t + \gamma)}+d_v) +c_u\omega \cos \omega t + v_{ext}.
	\label{project_V1_osc}
\end{eqnarray}
Eqs. (\ref{project_V0_osc}-\ref{project_V1_osc}) can be re-written as,
\begin{eqnarray}
  c_u\omega \cos{\omega t} +\frac{\alpha }{\tau A_u}(c_u\sin{\omega t}+d_u) =  \frac{A_v}{\tau A_u}c_v\sin{(\omega t + \gamma)}+\frac{A_v}{\tau A_u}d_v-v_{ext},
	\label{c1}\\
		 \frac{\alpha }{\tau A_u}(c_u\sin{\omega t}+d_u)  = \left(\frac{A_v}{\tau A_u}-\frac{mA_u}{\tau_v A_v}\right)(c_v\sin{(\omega t + \gamma)}+d_v) - c_v\omega \cos{(\omega t+\gamma)}.
	\label{c2}
\end{eqnarray}
At first glance, it may appear to be impossible to solve for six unknown variables ($c_u, c_v, \omega, \gamma, d_u$, and $d_v$) with only two equations. However, by applying the trigonometric transformation formula and $A_v = mA_u$, we can transform Eqs.~(\ref{c1}-\ref{c2}) into a form where both sides of two equations can be represented as sine waves, which are:
\begin{eqnarray}
  c_u\omega\sqrt{1+\left(\frac{\alpha}{\tau A_u \omega}\right)^2} \sin{(\omega t+\psi_1)} +\frac{\alpha }{\tau A_u}d_u &=&  \frac{m}{\tau}c_v\sin{(\omega t + \gamma)}+\frac{m}{\tau}d_v-v_{ext},
	\label{trans_1}\\
		c_v\sqrt{\tau_v^2\omega^2+1}\sin{(\omega t + \gamma+\psi_2)}+d_v &=& \tau_v \left(c_u\omega \cos \omega t + v_{ext}\right),
	\label{trans_2}
\end{eqnarray}
where
$\psi_1 =\arcsin (1/\sqrt{1+(\frac{\alpha}{\tau A_u \omega})^2})$,
and
$\psi_2 = \arcsin (\tau_v \omega/\sqrt{\tau_v^2\omega^2+1})$.

We can obtain six equations by equating the amplitudes, phase, and constant offset of both sides of Eqs.~(\ref{trans_1}, \ref{trans_2}), which are:
\begin{eqnarray}
  c_u\omega\sqrt{1+\left(\frac{\alpha}{\tau A_u \omega}\right)^2}  &=&  \frac{m}{\tau}c_v, 
	\label{e1}\\
  \psi_1 &=& \gamma, 
	\label{e2} \\ 
  \frac{\alpha}{\tau A_u}d_u &=& \frac{m}{\tau}d_v -v_{ext}, 
	\label{e3} \\ 
  c_v\sqrt{\tau_v^2\omega^2+1} &=& \tau_vc_u\omega,
	\label{e4} \\ 
  \gamma + \psi_2 &=& \frac{\pi}{2}, 
	\label{e5} \\ 
  d_v &=& \tau_v v_{ext}. 
	\label{e6} 
\end{eqnarray}
Finally, combining Eqs.(\ref{e1}-\ref{e6}), we solve the oscillatory tracking state, which are given by, 
	  \begin{eqnarray}
		  d_u &=& \frac{\tau_vA_u}{\alpha}v_{ext}(m-\frac{\tau}{\tau_v}),\\
		  d_v &=& v_{ext}\tau_v,\\
            \omega &=& \sqrt{ \frac{ 2\sqrt{\pi}\alpha ak(1+m)}{\tau \tau_v  (J_0+2\sqrt{\pi}ak\alpha)}},
            \label{eq-omega}\\
            \psi_1 &=& \arcsin (\frac{1}{\sqrt{1+(\frac{\alpha}{\tau A_u \omega})^2}}),\\
            c_u & = & c_v \sqrt{1+\frac{\tau A_u}{\tau_v \alpha}}.
	\end{eqnarray}
From Eq.~(\ref{eq-omega}), we see that
the oscillation frequency $\omega$ of the bump increases sublinearly with both the external input strength $\alpha$ and the adaptation strength $m$. Simulation results confirm these theoretical results (Fig.~\ref{fig4}d and e). 


\subsubsection{Stability analysis of the oscillatory tracking state}
We proceed to analyze the stability of the oscillatory tracking state.
As the adaptation strength $m$ continuously increase, the real parts of eigenvalues $\lambda_1$ and $\lambda_2$ keep increasing and their imaginary parts decrease to zero. The conditions for both eigenvalues being positive numbers are
\begin{eqnarray}
    \lambda_1+\lambda_2&=&1-\theta-\phi>0,
    \label{Real_part}\\
    \Delta^2&=&(\phi+\theta-1)^2-4\theta\phi\geq 0,
    \label{Imaginary_part}
\end{eqnarray}
which is equivalent to
\begin{eqnarray}
    \theta+\phi+\sqrt{\theta\phi} <1.
    \label{Osc_condition}
\end{eqnarray}
We consider that the external input speed $v_{ext}\ll a/\tau_v$ and the discrepancies $s_u\ll a$ and $s_v\ll a$. Under these conditions, we have the approximations $F(s_u) \approx F(s_v) \approx 1$ and $A_v \approx mA_u$ (see Eq.~(\ref{project_U1_tracking})). 
Substituting Eqs.~(\ref{theta}-\ref{phi}) and (\ref{project_U1_tracking}) into Eq.~(\ref{Osc_condition}), we get (?),
\begin{equation}
    m-\frac{\tau}{\tau_v} > \frac{\alpha}{A_u} \left(1 +\sqrt{\frac{\tau}{\tau_v}\frac{A_u}{\alpha}}\right). 
    \label{Osc_condition_approx}
\end{equation}
In this case, since both two eigenvalues are positive, the network state deviates from a fixed point, and the network falls into the state of travelling wave.

\subsubsection{Biological implications of oscillatory tracking}

We can intuitively understand how the A-CANN generates oscillatory tracking. Adaptation induces intrinsic mobility of the neural bump. Given that the strength of the external input is fixed. If the adaptation is too strong, the neural bump will move spontaneously as if no external input, realizing travelling wave; if the adaptation is too weak, the external input will drive the neural bump to move at its speed, realizing smooth tracking. Oscillatory tracking occurs when the adaptation strength has an intermediate value, whereby the neural bump tries to run away but is pulled back by the external input, causing the neural bump moves back and forth around the external input.  

Oscillatory tracking of the A-CANN can account for a number of theta response characteristics of hippocampal place cells during navigation found in rat experiments.
By choosing the parameters properly, we can set the frequency of bump oscillation in the range of theta band, independent of the speed of the external input. At the population level,
the oscillation of the bump center around the external input resembles theta sweeps of the decoded position based on place cell activities around the true position of the moving rat.
At the single neuron level, the forward and backward sweeps of the neural bump lead to that each neuron generates bursting responses at the theta rhythm, and the firing phases of a neuron in each theta cycle preceeds or proceeds, respectively, as the external input transfers through the receptive field of the neuron. This resembles the phase precession and procession phenomena observed in the expriments. For details, please refer to ~\cite{chu2022oscillatory}. Both theta sweeps and phase shift have been suggested to play important roles in spatial information encoding and memory formation.  

\subsection{Phase diagram of the tracking dynamics of the A-CANN}
From Eqs.~(\ref{Stationary_condition_approx},\ref{Osc_condition_approx}),  
we get the phase diagram of the tracking dynamics of A-CANN.
Particularly, by gradually increasing the adaptation strength $m$
while keeping
the external input strength $\alpha$ fixed (or equivalently, 
by keeping the adaptation strength $m$ fixed while gradually 
increasing
the external input strength $\alpha$), we have
\begin{itemize}
    \item \textbf{Smooth tracking state}, in which the neural bump
    tracks the external input at the speed $v_{ext}$ in a phase-locking manner. This occurs for  $m-\tau/\tau_v <\alpha/A_u$. Furthermore, depending on the relative amplitudes of $v_{ext}$ and $v_{int}$, the neural bump can either lead the external input when
    $v_{int}>v_{ext}$ or lag behind the input when $v_{int}<v_{ext}$.
    In particular, for small speed of the external input 
    ($v_{ext}\ll a/\tau_v$) and $m>\tau/\tau_v$, the neural bump tracks the external input with a constant anticipation time.

    \item \textbf{Oscillatory tracking state}, in which the neural bump tracks the external input in an oscillatory manner. This occurs for $\alpha/A_u<m-\tau/\tau_v<\alpha/A_u \left(1 +\sqrt{\frac{\tau}{\tau_v}\frac{A_u}{\alpha}}\right)$. 
    
    \item \textbf{Traveling wave state}, in which the neural bump moves freely at the speed $v_{int}$. This occurs for $m-\tau/\tau_v>\alpha/A_u \left(1 +\sqrt{\frac{\tau}{\tau_v}\frac{A_u}{\alpha}}\right)$. 
\end{itemize}
The phase diagram of the A-CANN is presented in Fig.~\ref{fig4}f. Our theoretical analyses agree well with simulation results.

\section{Noisy adaptation generates Diffusion and Super-diffusion Dynamics in CANNs}

In the previous sections, we have discussed the dynamics of the A-CANN without noise. In reality, however, real biological nervous systems are full of endogenous and exogenous noises at neuronal dynamics and signal transmission. Extensive researches have shown that noises introduce randomness in network dynamics, which can enhance neural information processing (see, e.g., \cite{orban2016neural,qi2022fractional,haefner2016perceptual}). 
In the below, we study how noises affect the dynamics of the A-CANN and their implications on neural information processing. In particular, in addition to noises on the neuronal dynamics, we also consider noises on the adaptation dynamics. 

With noises, the dynamics of the A-CANN 
are re-written as,
\begin{eqnarray}	
\tau \frac{dU(x,t)}{dt} & = & -U(x,t)+\rho \int_{-\infty}^{\infty} J(x,x')r(x',t)\, dx'-V(x,t) + I^{ext}(x,t) +\sigma_U\xi_U(x,t), \nonumber \\
\label{dynamics-U-noisy}  
\\
\tau_v \frac{dV(x,t)}{dt} &=& -V(x,t)+mU(x,t)+\sigma_m\xi_m(x,t)f\left[U(x,t)\right], 
\label{dynamics-V-noisy}
\end{eqnarray}
where the term $\sigma_U\xi_U(x,t)$ represents noises on the neuronal dynamics, with $\sigma_U$ the noise strength and $\xi_U(x,t)$ Gaussian white noise of zero mean and unit variance. The term $\sigma_m\xi_m(x,t)f[U(x,t)]$ represents noises on the adaptation dynamics, with $\sigma_m$ controlling the noise strength and $\xi_m(x,t)$ Gaussian white noise of zero mean and unit variance; the function $f[U(x,t)]$ denoting how the noise strength depends on the neuronal activity. 

\subsection{Spontaneous dynamics of the noisy A-CANN}

We first study the spontaneous dynamics of the noisy A-CANN by setting $I^{ext}(x,t) = 0$. We consider $f[U(x,t)]= U(x,t)$ for the clarity of theoretical analysis. Again, we assume that the network state can be approximately to be of the Gaussian form, which hold when noises are not too strong. 

Substituting Eqs.~(\ref{Ubar}-\ref{Vbar}) into Eqs.~(\ref{dynamics-U-noisy}-\ref{dynamics-V-noisy}) and (\ref{dynamics-r}), we obtain,
\begin{eqnarray}
  A_r&=&\frac{A_u^2}{1+k\rho\sqrt{2\pi}aA_u^2},
  \\
  \tau\left[A_u \frac{x-z}{2a^2}\frac{dz}{dt}+\frac{dA_u}{dt}\right]\mathcal{N}(z,2a) & = & 
    (-A_u+\frac{\rho J_0}{\sqrt{2}}A_r)\mathcal{N}(z,2a)\nonumber \\&& 
    -A_v\mathcal{N}(z-s,2a)
    +\sigma_U\xi_U(x,t),  \nonumber \\
    \label{dynamics_u(t)_noisy}
    \\
    \tau_v\left[A_v \frac{x-z+s}{2a^2}\frac{d(z-s)}{dt}+\frac{dA_v}{dt}\right] \mathcal{N}(z-s,2a)
	 & = & -A_v\mathcal{N}(z-s,2a) \nonumber \\ && 
    +[m+\sigma_m\xi_m(x,t)]A_u\mathcal{N}(z,2a), \nonumber \\
    \label{dynamics_v(t)_noisy}
\end{eqnarray}
where $s$ denotes the separation between $U$ and $V$ bumps, while other notations keep the same as in previous sections. 

We project Eqs.~(\ref{dynamics_u(t)_noisy}-\ref{dynamics_v(t)_noisy}) onto the two dominating motion modes of the network dynamics, i.e., the bump height ($u_0(x|z)$) and position ($u_1(x|z)$), and obtain,
  \begin{eqnarray}
    \tau \frac{dA_u}{dt} & = &  (-A_u+\frac{\rho J_0}{\sqrt{2}}A_r) - A_v\exp \left(-\frac{s(t)^2}{8a^2}\right) +\frac{1}{\sqrt{2\pi}a}\sigma_U\xi_{U,0}(t), 
    \label{project_U0_noisy}\\
    \tau A_u \frac{dz}{dt} & = & s(t) A_v\exp\left(-\frac{s(t)^2}{8a^2}\right)+\sqrt{\frac{2}{\pi}}\sigma_U\xi_{U,1}(t),
    \label{project_U1_noisy} \\
     \tau_v \frac{dA_v}{dt} &=& -A_v +[m+\frac{1}{2\sqrt{\pi}a}\sigma_m\xi_{m,0}(t)]A_u \exp\left(-\frac{s(t)^2}{8a^2}\right),
    \label{project_V0_noisy} \\
    \tau_v A_v \frac{d s}{dt} & = &  s(t) \exp\left(-\frac{s(t)^2}{8a^2}\right)\left(\frac{\tau_v}{\tau A_u}A_v^2 - [m+\frac{1}{2\sqrt{\pi}a}\sigma_m\xi_{m,0}(t)]A_u \right)\nonumber \\ &&
    +\frac{\tau_v A_v}{\tau A_u}\sqrt{\frac{2}{\pi}}\sigma_m\xi_{m,1}(t),
    \label{project_V1_noisy} 
  \end{eqnarray}
where $\xi_{U,0}(t)$, $\xi_{U,1}(t)$, $\xi_{m,0}(t)$, $\xi_{m,1}(t)$ are the projections of $\xi_{U}(t)$ and $\xi_{m}(t)$ onto the two motion modes, respectively, and they are still Gaussian noises of zero mean and unit variance.

The previous sections have shown that 
in the case of no noises, $m_0=\tau/\tau_v$ is the critical point, below which ($m<m_0$) the network holds the static bump state and above which ($m>m_0$) the network holds the travelling wave state. Intuitively, it is understandable that when Gaussian white noises are included, they causes the neural bump to experience local Brownian motion when $m\ll m_0$, while they have little effect on
the travelling wave when $m\gg m_0$. The interesting phenomenon happens when $m$ is close to the boundary $m_0$, as noises will cause the effective adaptation strength $\tilde{m}(t)$ to fluctuate around the boundary, leading to that the neural bump to 
exhibit a mixed movement pattern: temporally Brownian motion when $\tilde{m}(t)<m_0$ and temporally large-size movement caused by travelling wave when $\tilde{m}(t)<m_0$. Such a motion pattern is called Lévy flights. In below, we analyze this Lévy flights state. 

For the convenience of analysis, we consider that the levels of noises are low and that the variances of bump heights $A_u$ and $A_v$ are negligible. Under these approximations, Eqs.~(\ref{project_U1_noisy},\ref{project_V1_noisy}) can be simplified as,
  \begin{eqnarray}
    \tau\frac{dz}{dt} & = & ms(t)+\sqrt{\frac{2}{\pi}}\frac{\sigma_U}{A_u}\xi_{U,1}(t),
    \label{project_U1_noi} \\
    \tau_v \frac{d s}{dt} & = &  -[1-\frac{\tau_v}{\tau}m+\frac{\sigma_m}{2\sqrt{\pi}a m}\xi_{m,0}(t)]s(t)
    +\sqrt{\frac{2}{\pi}(\frac{\tau_v\sigma_U}{\tau A_u})^2 + \frac{1}{2\pi}(\frac{\sigma_m}{m})^2}\xi_s(t), \nonumber \\
    \label{project_V1_noi} 
  \end{eqnarray}
where $\xi_s(t)$ is Gaussian white noise of zero mean and unit variance. 

Eq.~(\ref{project_U1_noi}) shows that the bump position $z(t)$ is determined by two parts, a drift term ($m s(t)$) reflecting the contribution of adaptation and a diffusion term ($\sqrt{2/\pi}(\sigma_U/A_u)\xi_{U,1}(t)$) reflecting the contribution of noises on the neuronal dynamics. 
It is evident that, in the absence of adaptation ($m=0$), the dynamics of the bump center expressed by Eq.~(\ref{project_U1_noi}) degenerate into a Brownian motion process driven by Gaussian noise~\cite{bartumeus2007Levy}.

To solve the dynamics of $z(t)$, we first solve the dynamics of $s(t)$. For clearance, we re-organize Eq.~(\ref{project_V1_noi}) as, 
\begin{eqnarray}
\tau_v \frac{ds}{dt}=-(\mu+\gamma\xi_{m,0})s+\sigma_s\xi_s, \label{dynamics_s_noisy}
\end{eqnarray}
where $\mu=1-m\tau_v/\tau$ denotes the distance-to-boundary ratio, reflecting the gap between the mean adaptation strength and the boundary of the travelling wave ($\tau/\tau_v$). $\gamma=\sigma_m/(2\sqrt{\pi}am)$ denotes the noise-to-strength ratio, reflecting the relative amplitude of noises. $\sigma_s=\sqrt{2\tau_v^2\sigma_U^2/(\pi \tau^2 A_u^2)+2a^2\gamma^2}$ represents the re-scaled neuronal noises.

We first consider that no noise in the adaptation ($\gamma=0$). In such a case, if $\mu<0$, i.e., $m>\tau/\tau_v$, $s(t)$ will simply increase monotonically with time without converging to any stable points. 
If $\mu>0$, i.e., $0<m<\tau/\tau_v$, the evolution of the equation proposed in Eq.~(\ref{dynamics_s_noisy}) gives rise to an Ornstein-Uhlenbeck (OU) process~\cite{uhlenbeck1930theory}, and the distribution of $s(t)$ is Gaussian~\cite{uhlenbeck1930theory}. It can be written as,
\begin{eqnarray}
p^{st}(s)=\sqrt{\frac{\mu}{\pi\sigma_s^2}}\exp\left[-\frac{\mu s^2}.{\sigma_s^2}\right]
\label{s_distribution_Ou}
\end{eqnarray}
Combining Eqs.~(\ref{s_distribution_Ou}) and (\ref{project_U1_noi}), the dynamics of $dz/dt$ turns out to be determined by the sum of two Gaussian noise terms, which is still Gaussian. This implies that the bump takes on Brownian motion.

The dynamics of the bump center $z(t)$ becomes intriguing when the adaptation strength $m$ fluctuates, i.e. $\gamma>0$, since the multiplicative noise $\xi_m$ in the drift term will play its role. Utilizing Ito calculus, we can rewrite Eq.~(\ref{dynamics_s_noisy}) as a first order difference equation,
\begin{eqnarray}
s(t+dt) & = & s(t)+\int_t^{t+dt}\left[
-\frac{\mu s(t')}{\tau_v}+\frac{\gamma s(t')}{\sqrt{\tau_v}}\xi_m(t')+\frac{\sigma_s}{\sqrt{\tau_v}}\xi_s(t')\right] dt',\nonumber \\
&&
= s(t)-\frac{\mu s(t)}{\tau_v}dt+\frac{1}{\sqrt{\tau_v}}\left[-s(t)\gamma dt\bar{\xi_m}+\sigma_sdt\bar{\xi_s}\right],
\label{s_first_order}
\end{eqnarray}
where $dt\bar{\xi_m}$ and $dt\bar{\xi_s}$ are the Ito prescriptions in the limit of $dt\xrightarrow{}0$.
To derive the Fokker-Planck equation, we adopt a smooth trail function $R(s)$ proposed by Rivers\cite{rivers1988path} and calculate its time average,
\begin{eqnarray}
\left \langle R(t) \right \rangle = \int R(s)p(s,t)ds,
\end{eqnarray}
where $p(s,t)$ is the distribution of $s(t)$ at time $t$. The evolution of the average value of $R(t)$ in a short interval $dt$ at time $t$ is given by,
\begin{eqnarray}
\left \langle R(t+dt) \right \rangle & = &  \left \langle \int R\left(s-\frac{\mu s}{\tau_v}dt+\frac{1}{\sqrt{\tau_v}}(-s\gamma dt\bar{\xi_m}+\sigma_sdt\bar{\xi_s})\right) p(s,t)ds\right \rangle,\\
& = &  \left \langle\int \left(R(s)+dtR'(s)(-\frac{\mu}{\tau_v}s)+dtR''(s)(\frac{\sigma_s^2+\gamma^2s^2}{2\tau_v})\right) p(s,t)ds\right \rangle . \nonumber \\
\label{Taylor_R}
\end{eqnarray}
Eq.~(\ref{Taylor_R}) is the first-order Taylor-series expansion of the trail function $R(\cdot)$. 
Note that the left side corresponds to the partial derivative of $p(s,t)$ with respect to $t$, and the right side corresponds to the partial derivative of $p(s,t)$ with respect to $s$. Thus we obtain the Fokker-Planck equation which describes the time evolution of the distribution of $s(t)$,
\begin{eqnarray}
\frac{\partial p(s,t)}{\partial t}=-\frac{\partial}{\partial s}\left(-\frac{\mu}{\tau_v}sp(s,t)\right)+\frac{\partial^2}{\partial s^2}\left(\frac{\sigma_s^2+\gamma^2s^2}{2\tau_v}p(s,t)\right).
\end{eqnarray}
The stationary distribution of $p^{st}(s)$ is achieved when
\begin{eqnarray}
-\frac{\mu}{\tau_v}sp(s,t)=\frac{d}{ds}\left(\frac{\sigma_s^2+\gamma^2s^2}{2\tau_v}p(s,t)\right).
\end{eqnarray}
Thus we obtain the stationary distribution of $s(t)$ by solving the corresponding Fokker-Planck equation, which gives,
\begin{eqnarray}
p(s)=c_0(\sigma_s^2+\gamma^2s^2)^{-(1+\mu/\gamma^2)},
\label{s_distribution}
\end{eqnarray}
where $c_0$ is a normalization constant.

According to Eq.~(\ref{project_U1_noi}), the displacement of $z(t)$ in a short time interval $\delta t$ is calculated to be $||\Delta z|| = ||ms\delta t/\tau+\sqrt{2\delta t/(\pi\tau)}\sigma_U/A_u\xi_{U,1}|| $. Substituting $s$ with its stationary distribution Eq.~(\ref{s_distribution}), we finally derive the distribution of $||\Delta z||$,
\begin{eqnarray}
p(||\Delta z||)=c_1||\Delta z||^{-1-\alpha},
\label{z_distribution}
\end{eqnarray}
which clearly satisfies a power-law distribution with the exponent $\alpha = 1+2\mu/\gamma^2$.

\begin{figure}[H]
\hfill
\begin{center}
\includegraphics[width=15 cm]{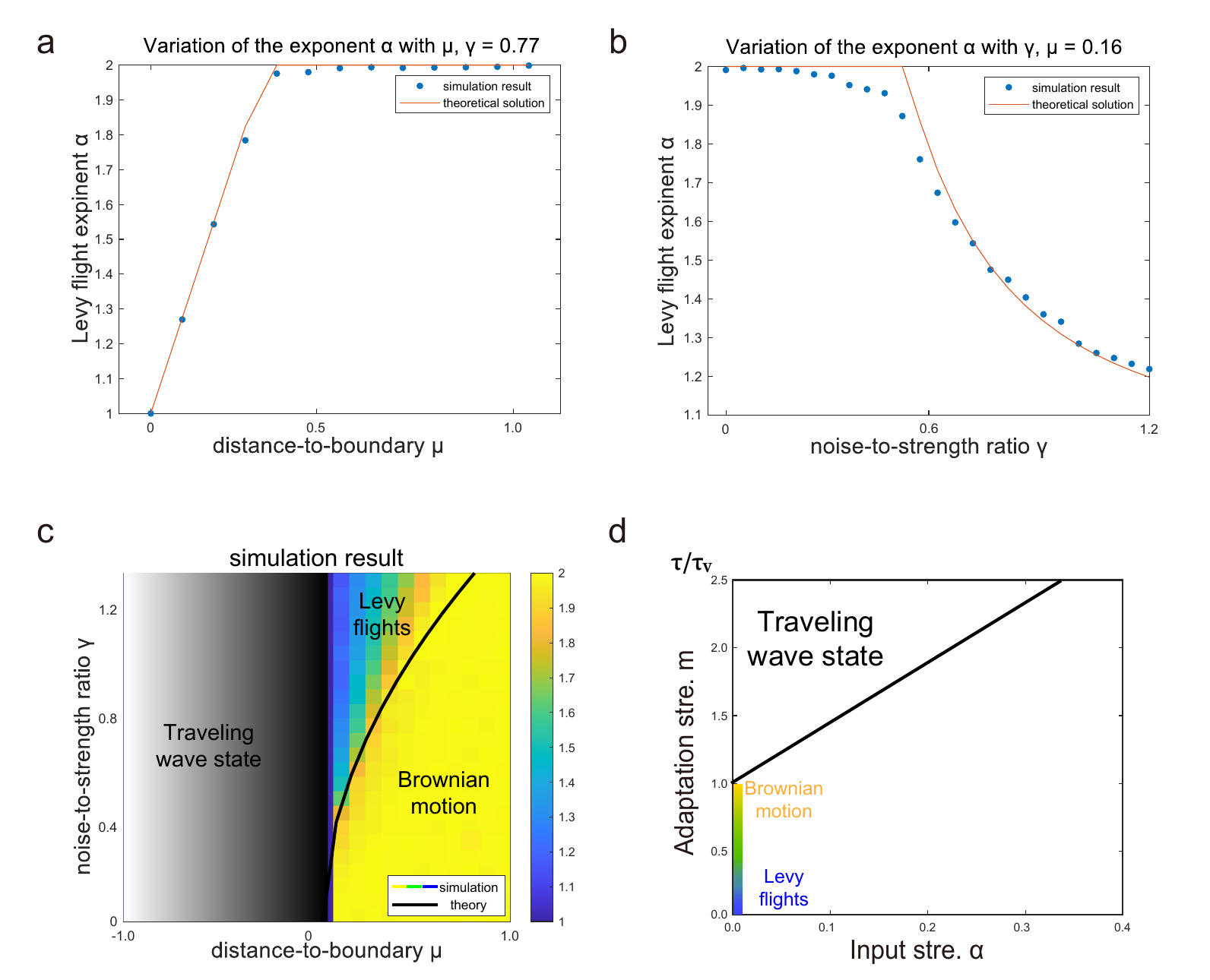}
\end{center}
\small
\caption{Spontaneous dynamics of the noisy A-CANN.
\textbf{a}. The Lévy exponent $\alpha$ vs. $\mu$ with $\gamma
 = 0.77$. Note that when $\mu>\gamma^2/2$, all  $\alpha > 2$ (Brownian motion) will converge to $\alpha = 2$ due to the Central Limit Theorem.
\textbf{b}. The Lévy exponent $\alpha$ vs. $\gamma$ with $\mu
 = 0.16$.
\textbf{c}. The phase diagram of the network dynamics with respect to the distance-to-boundary $\mu$ and the noise-to-strength ratio $\gamma$.
\textbf{d}. The phase diagram of the network dynamics with respect to the input strength $\alpha$ and the adaptation strength $m$.
Parameters are: $a = \pi/10$, $J_0 = 10$, $k = 0.05$,
$\tau= 10ms$,$\tau_v= 25ms$, $\rho = 1600/(2\pi)$, $\bar{m} = 0.56$, $\sigma_m = 6.4$, $\sigma_U=0.01$.
}\label{fig5}
\end{figure}

The above power-law distribution represents a typical diffusion (or random walk) process. When the diffusion parameter $\alpha$ satisfies $0<\alpha<2$, the distribution describes Lévy flights, also termed as anomalous diffusion or super diffusion. When $\alpha>2$, the above process degenerates to Brownian motion due to the Central Limit Theorem~\cite{bartumeus2007Levy}. Although both are random wanderings, Lévy flight and Brownian motion have different properties. The distribution of steps in Brownian motion can be drawn as a bell-shaped curve. Lévy flights, on the other hand, are a class of random processes with Markovian properties characterized by occasional long-range jumps, i.e. long-tailed asymptotic form as $|s|\xrightarrow{}\infty$ . Therefore, Lévy flights are more prone to occasional long-range jumps in comparison to Brownian motion. 
We carry simulations to verify our theoretical results. Fig.~\ref{fig5}c  shows the phase diagram of the spontaneous dynamics of noisy A-CANN. Fig.~\ref{fig5}a and b shows the comparison between the simulation results and the theoretical solutions of the distribution exponent $\alpha$.

\subsubsection{Biological implications of Lévy flights}

The A-CANN with adaptation has the intrinsic mobility of generating a travelling wave when the adaptation strength is strong enough~\cite{bressloff2011spatiotemporal,mi2014spike}. The condition of $m_0 =\tau/\tau_v$ defines the boundary of the traveling wave. Above the boundary, the bump moves spontaneously in the attractor space; while below the boundary, the bump remains static without external input or exhibits Brownian motion when external noise is present. 
When noises are considered in the adaptation strength, the A-CANN displays new interesting dynamical behaviors. 
Specifically, consider the mean of the adaptation strength is close to the traveling wave boundary, noises occasionally push the adaptation strength across the boundary, causing bump movement to temporarily fall into the traveling wave state, and the bump travels a long distance (long jump) in the attractor space. Over time, due to the adaptation strength fluctuations, the bump movement displays intermittent local Brownian motion (the adaptation strength below the boundary) and long-jump motion (the adaptation strength above the boundary) , manifesting the characteristic of Lévy flights~\cite{dong2021noisy}.

Mathematically, Lévy flight is an efficient strategy for information search in an unknown open space~\cite{viswanathan1996Levy,bartumeus2005animal, reynolds2009scale}.
It has been widely found in animal forging behaviors and also
in human psychology experiments, such as free memory recalls~\cite{rhodes2007human, costa2016foraging} .
Furthermore, recently, the data recorded in electrophysiological experiments also revealed that when experimental animals remain still, the sequential reactivation of place cells in the hippocampal region follow statistical characteristics of Lévy flights~\cite{pfeiffer2015autoassociative}. Our model may offer a potential mechanism for explaining the patterns of sequential activation of place cells, thus providing more insights into the understanding of higher cognitive functions associated with them.

\unskip

\section{Conclusion and Discussion}

Attractor networks, such as the Hopfield network, are neural models where information is encoded as stationary states in a dynamic system of interconnected neurons. While the Hopfield model effectively explains associative memory with discrete attractors, it falls short for correlated or continuous information representations. This led to the development of continuous attractor neural networks (CANNs), which form continuous subspaces for smooth tracking of information like orientation or spatial location. By studying CANNs, we gain insights into the balance between stability and mobility in neural representations, crucial for both reliable information encoding and efficient processing. Adaptive mechanisms, like neural adaptation, introduce necessary dynamics for this balance. In this research, ‘neural adaptation’ refers to the common phenomenon of decaying neuronal activities in response to repeated or prolonged stimulation in neural networks. 

Here we review and unify previous studies on adaptive CANNs, offering a comprehensive theoretical framework and discussing the resulting rich dynamical behaviors and their biological implications.

Although the equations governing the dynamics of continuous attractor neural networks (CANNs) seem complex due to their high dimensionality, their behavior is primarily influenced by a few dominant motion modes, such as changes in the bump's height, position, and width (illustrated in Fig.2d). By focusing on these dominant modes, we can greatly simplify the analysis of the network behavior by turning the high-dimensional neurodynamic model into state space model.

The role of adaptation mainly manifests in enhancing the intrinsic mobility and expressing capabilities of the system. In summary, external input pattern and intrinsic parametric conditions together determine the dynamical behavior of the system. 

Considering spontaneuous motion (without external input), A-CANN holds a static bump as its stationary state when both the inhibition and adaptation are sufficiently small, satisfying $0<k<k_{c2}$ and $m<\tau/\tau_v$. These agree with our intuition, as strong inhibition will suppress neuronal activities. If $m>\tau/\tau_v$, the strong adaptation will drag and untimately destabilize the static bump,
causing the bump to move spontaneously in the feature space in a traveling wave manner. The wave enables a neural system to progressively visit all stationary points. 
This could endows cognitive functions of memory retrieval and memory consolidation ~\cite{carr2011hippocampal}. 

In a practical situation, the neural system receives external or upstream input all the time. If the adaptation is weak, the external input will drive the neural bump to move at its speed, realizing smooth tracking; if the adaptation is too strong, the neural bump will move spontaneously as if no external input, realizing previous travelling wave. When taking an intermediate value, oscillatory tracking occurs, in which the neural bump is attracted by the center of the external input and moves back and forth. Our bifurcation analysis explains the state transition principle of the aattractor neural network from the perspective of dynamic system, and the theoretical and simulation results fit well (Fig.\ref{fig2}).

When noise is factored, the A-CANN exhibits novel dynamical behaviors. Specifically, when the mean adaptation strength approaches the boundary value of the traveling wave, noise perturbations occasionally push the adaptation strength beyond the boundary. This causes the bump movement to switch between intermittent local Brownian motion and long-jump motion, demonstrating the characteristic of Lévy flights~\cite{dong2021noisy}. This behavior enables efficient searching in neural landscape and can function as a possible policy in information processing.

By responding to external stimuli and processing internal information, adaptation mechanisms enable systems to continuously adjust their structure and parameters to adapt to environmental changes and emerge more diverse dynamical states. This dynamic adjustment ability is crucial for the survival and evolution of complex systems in complicated changing environments. The manifestation of adaptation in external phenomena is reflected in the enhancement of the stability, robustness, and self-organization of the system. Through adaptation mechanisms, systems can maintain stable states under external disturbances and possess a certain self-repair capability. The computational significance of adaptation mechanisms is self-evident. 

The implementation foundation of adaptation mechanisms lies in the sensitivity to dynamically adjust information processing according to current state. In artificial neural networks, this is typically achieved through weight adjustments, gain modulation, etc., while in biological neural networks, it involves spiking frequency adaptation, synaptic plasticity, and neuron alterations. Here we summarize the possible biological applications of the dynamical properties of A-CANN in the various states mentioned earlier.

Adaptation causes suppression around network bump, destabilizing it and driving it to move. This spontaneous movement, aided by recurrent neural interactions, results in traveling wave behavior. Such waves could facilitate active retrieval of stored memories, supporting cognitive functions like memory search and consolidation~\cite{luhmann2016spontaneous, mitra2016networks, carr2011hippocampal}.

CANNs track external moving inputs but typically lag behind. However, with strong adaptation, A-CANNs can track inputs anticipatively. The adaptation creates intrinsic mobility, causing the neural bump to move at an internal speed, competing with the speed of the external input. When the internal speed exceeds the external speed, the bump leads the input with a constant time. This mechanism could help compensate for neural signal transmission delays, ensuring timely perception and reaction, as seen in the head-direction system of rats~\cite{blair1995anticipatory}.

A-CANNs exhibit oscillatory tracking when adaptation strength is intermediate. This may theta oscillations in hippocampal place cells during navigation, where the bump's oscillation frequency matches the theta band. This results in theta sweeps and phase shifts in neuron firing, contributing to spatial information encoding and memory formation~\cite{chu2022oscillatory}.

In the A-CANN, adaptation can cause the neural bump to move spontaneously, generating traveling waves when adaptation strength is high. Near the threshold of this strength, noise can push the system into traveling wave states, resulting in intermittent long jumps and local Brownian motion, characteristic of Lévy flights~\cite{dong2021noisy}. Lévy flights are efficient for information search and have been observed in animal foraging and human memory recall. This model may explain the sequential activation patterns of hippocampal place cells during rest, providing insights into higher cognitive functions.

Overall, our study explores various dynamic behaviors of A-CANNs and their biological implications. The traveling wave behavior supports memory retrieval and cognitive functions. Anticipative tracking compensates for neural delays. Oscillatory tracking explains hippocampal theta rhythms. Lévy flights offer an efficient mechanism for information search and hippocampal place cell activation patterns. These findings provide a comprehensive understanding of how neural adaptation influences brain functions and cognition.

We highlights the key differences between A-CANNs and typical CANNs, emphasizing the crucial role of adaptation. Neural adaptation mechanisms introduce dynamic flexibility to the otherwise stable attractor states. This extra flexibility allows A-CANNs to explain a broader range of experimental phenomena, such as traveling waves, anticipative tracking, oscillatory behavior, and Lévy flights, observed in various neural systems.

The ability of our network to align with these diverse neural encoding phenomena suggests that A-CANN may represent a general mechanism for neural information storage and processing. Specifically, neural system may combine attractor dynamics with adaptation, effectively balancing stability and flexibility. Attractor dynamics ensure stable information storage, while adaptation facilitates flexible information search and retrieval. This dynamic interplay between stability and mobility enables the neural system to maintain robust representations while efficiently adapting to new experience and environmental changes.

In conclusion, adaptation mechanisms play an important and far-reaching role in abounding the dynamical properties of standard CANNs, with profound implications. This review not only contributes to a deeper understanding of the CANN but also provides new ideas and methods for the development and application of adaptation mechanism in neural networks under a unified analysis framework. It is hoped that the research in this paper can provide some reference and inspiration for further exploration and development in related fields.



%

\subsection*{Acknowledgments}
This work was support by: Science and Technology Innovation 2030-Brain Science and Brain-inspired Intelligence Project (No. 2021ZD0200204, SW).

\section*{Appendix A: Gaussian bump}\label{Appendix A}

Considering the translational-invariant connection
\begin{eqnarray}
J(x,x')=\frac{J_0}{\sqrt{2\pi}a}\exp[-(x-x')^2/(2a^2)]
\end{eqnarray}
and the Gaussian shape approximation of the bump
\begin{eqnarray}	
	U_0(x,t) & = & A_u \exp \left\{-\frac{\left[x-z(t)\right]^2}{4a^2}\right\}, 
	\label{Ubar_a}\\
	V_0(x,t) & = & A_v \exp \left\{-\frac{\left[x-z(t)+s_v(t)\right]^2}{4a^2}\right\}, 
	\label{Vbar_a}\\
	r_0(x,t) & = & A_r \exp \left\{-\frac{\left[x-z(t)\right]^2}{2a^2}\right\},
	\label{rbar_a}
\end{eqnarray}
then the network dynamics
\begin{eqnarray}	
\tau \frac{dU(x,t)}{dt} & = & -U(x,t)+\rho \int_{-\infty}^{\infty} J(x,x')r(x',t)\, dx'-V(x,t)+I^{ext}(x,t), 
\label{dynamics-U_a} 
\\
r(x,t) &= &\frac{U(x,t)^2}{1+k\rho \int_{-\infty}^{\infty} U^2(x',t), dx'},
\label{dynamics-r_a}
\\
\tau_v \frac{dV(x,t)}{dt} &= &-V(x,t)+mU(x,t), \label{dynamics-V_a}
\end{eqnarray}
can be simplified as

\begin{eqnarray}
  A_r&=&\frac{A_u^2}{1+k\rho\sqrt{2\pi}aA_u^2},
  \label{dynamics_r_eqn_a}\\
  \tau\left[A_u \frac{x-z}{2a^2}\frac{dz}{dt}+\frac{dA_u}{dt}\right]exp[-\frac{(x-z)^2}{4a^2}] & = & 
    (-A_u+\frac{\rho J_0}{\sqrt{2}}A_r)exp[-\frac{(x-z)^2}{4a^2}]\nonumber \\ && -A_v exp[-\frac{(x-z+s_v)^2}{4a^2}]+I^{ext}(x,t),\label{dynamics_u(t)_eqn_a} \nonumber
    \\
    \\
    \tau_v\left[A_v \frac{x-z+s_v}{2a^2}\frac{d(z-s_v)}{dt}+\frac{dA_v}{dt}\right] exp[-\frac{(x-z+s_v)^2}{4a^2}]
	 & = & -A_vexp[-\frac{(x-z+s_v)^2}{4a^2}] \nonumber \\
	 && +mA_uexp[-\frac{(x-z)^2}{4a^2}], 
    \label{dynamics_v(t)_eqn_aa}
\end{eqnarray}

\section*{Appendix B: projection method}\label{Appendix B}

For the bump $U(x,t)$, the first two motion modes are,
\begin{eqnarray}	
		  u_0(x,t) & = & \exp \left\{-\frac{[x-z(t)]^2}{4a^2}\right\},
		  \\
		  u_1(x,t) & = & \left[x-z(t)\right]\exp \left\{-\frac{[x-z(t)]^2}{4a^2}\right\}.
\end{eqnarray}
For the bump $V(x,t)$, the first two motion modes are,
\begin{eqnarray}	
	v_0(x,t) & = & \exp \left\{-\frac{[x-z(t)+d(t)]^2}{4a^2}\right\}, \\
	v_1(x,t) & = & \left[x-z(t)+d(t)\right]\exp \left\{-\frac{[x-z(t)+d(t)]^2}{4a^2}\right\}.
\end{eqnarray} 
Projecting network dynamics Eqs.~(\ref{dynamics_u(t)_eqn}) onto the first two dominant motion modes of $u$ ($\int _{-\infty}^\infty f(x)u(x)dx$), we obtain

\begin{eqnarray}
  Left & = & \int\tau \left[ A_u  \frac{x-z}{2a^2}\frac{dz}{dt}+\frac{dA_u}{dt}\right]exp\left\{-\frac{(x-z)^2}{2a^2}\right\}dx=\sqrt{2\pi} a \tau \frac{dA_u}{dt}, \\
  right & = & \int(-A_u+\frac{\rho J_0}{\sqrt{2}}A_r)exp\left\{-\frac{(x-z)^2}{2a^2}\right\}dx \nonumber  -\int A_v exp\left\{-\frac{(x-z+s_v)^2+(x-z)^2}{4a^2}\right\}dx \nonumber \\ && 
    +\int I^{ext}exp\left\{-\frac{(x-z)^2}{4a^2}\right\}dx , \nonumber
  \\ 
    &=&\sqrt{2\pi} a (-A_u+\frac{\rho J_0}{\sqrt{2}}A_r) - \sqrt{2\pi} a A_v\exp \left(-\frac{s_v(t)^2}{8a^2}\right) +\int  I^{ext}exp\left\{-\frac{(x-z)^2}{4a^2}\right\}dx,
\end{eqnarray}
for $u_0$, and
\begin{eqnarray}
  Left & = & \int\tau \left[ A_u \frac{(x-z)^2}{2a^2}\frac{dz}{dt}+(x-z)\frac{dA_u}{dt}\right]exp\left\{-\frac{(x-z)^2}{2a^2}\right\}dx , \nonumber \\
  &=&2a^2\tau A_u \frac{dz}{dt},\\
  right & = & \int(x-z)(-A_u+\frac{\rho J_0}{\sqrt{2}}A_r)exp\left\{-\frac{(x-z)^2}{2a^2}\right\}dx \nonumber \\ && -\int (x-z)A_v exp\left\{-\frac{(x-z+s_v)^2+(x-z)^2}{4a^2}\right\}dx  \nonumber \\ && 
    +\int (x-z)I^{ext}exp\left\{-\frac{(x-z)^2}{4a^2}\right\}dx , \nonumber
    \\
    & = &2a^2s_v A_v\exp\left(-\frac{s_v^2}{8a^2}\right)+\int (x-z)I^{ext}exp\left\{-\frac{(x-z)^2}{4a^2}\right\}dx ,
\end{eqnarray}
for $u_1$.

We can synthesize the above equations as
  \begin{eqnarray}
    \tau \frac{dA_u}{dt} & = &  (-A_u+\frac{\rho J_0}{\sqrt{2}}A_r) - A_v\exp \left(-\frac{s_v(t)^2}{8a^2}\right) +\int_{x} I^{ext}(x,t)u_0(x)dx, 
    \\
    \tau A_u \frac{dz}{dt} & = & s_v(t) A_v\exp\left(-\frac{s_v(t)^2}{8a^2}\right)+\int_{x} I^{ext}(x,t)u_1(x)dx . \label{z_a}
  \end{eqnarray}

Projecting network dynamics Eqs.~(\ref{dynamics_u(t)_eqn}) onto the first two dominant motion modes of $u$, we obtain
\begin{eqnarray}
    \tau_v\left[A_v \frac{x-z+s_v}{2a^2}\frac{d(z-s_v)}{dt}+\frac{dA_v}{dt}\right] exp[-\frac{(x-z+s_v)^2}{4a^2}]
	 & = & -A_vexp[-\frac{(x-z+s_v)^2}{4a^2}] \nonumber \\
	 && +mA_uexp[-\frac{(x-z)^2}{4a^2}], 
    \label{dynamics_v(t)_eqn_a}
\end{eqnarray}
\begin{eqnarray}
  Left & = & \int\tau_v\left[A_v \frac{x-z+s_v}{2a^2}\frac{d(z-s_v)}{dt}+\frac{dA_v}{dt}\right]exp\left\{-\frac{(x-z+s_v)^2+(x-z)^2}{4a^2}\right\}dx , \nonumber \\
  &=&\sqrt{2\pi}a\tau_v\left[A_v \frac{s_v}{4a^2}\frac{d(z-s_v)}{dt}+\frac{dA_v}{dt} \right]\exp \left(-\frac{s_v^2}{8a^2}\right) , \\
  right & = & \int m A_u exp\left\{-\frac{(x-z)^2}{2a^2}\right\}dx  -\int A_v exp\left\{-\frac{(x-z+s_v)^2+(x-z)^2}{4a^2}\right\}dx \nonumber \\ && 
    +\int I^{ext}exp\left\{-\frac{(x-z)^2}{4a^2}\right\}dx , \nonumber
  \\ 
    &=&\sqrt{2\pi}amA_u - \sqrt{2\pi}a A_v\exp \left(-\frac{s_v^2}{8a^2}\right) +\int  I^{ext}exp\left\{-\frac{(x-z)^2}{4a^2}\right\}dx ,
\end{eqnarray}
for $u_0$, and
\begin{eqnarray}
  Left & = & \int\tau_v\left[A_v \frac{x-z+s_v}{2a^2}\frac{d(z-s_v)}{dt}+\frac{dA_v}{dt}\right](x-z)exp\left\{-\frac{(x-z+s_v)^2+(x-z)^2}{4a^2}\right\}dx , \nonumber \\
  &=&\sqrt{2\pi}a\left[\frac{s_v}{2}\frac{dA_v}{dt}-(\frac{s_v^2}{8a} -\sqrt{\frac{\pi}{2}}a)A_v \frac{d(z-s_v)}{dt}\right]\exp \left(-\frac{s_v^2}{8a^2}\right) , \\
  right & = & \int m A_u (x-z)exp\left\{-\frac{(x-z)^2}{2a^2}\right\}dx  -\int A_v (x-z)exp\left\{-\frac{(x-z+s_v)^2+(x-z)^2}{4a^2}\right\}dx \nonumber \\ && 
    +\int (x-z)I^{ext}exp\left\{-\frac{(x-z)^2}{4a^2}\right\}dx , \nonumber
  \\ 
    &=&\sqrt{2\pi}a A_v\frac{s_v}{2}\exp \left(-\frac{s_v^2}{8a^2}\right) +\int (x-z) I^{ext}exp\left\{-\frac{(x-z)^2}{4a^2}\right\}dx ,
\end{eqnarray}
for $u_1$.

Combine these two equations and separate $A_v$ and $s_v$ using Eq.~(\ref{z_a}), we have
  \begin{eqnarray}
     \tau_v \frac{dA_v}{dt} &=& -A_v +mA_u \exp\left(-\frac{s_v(t)^2}{8a^2}\right),
    \\
        \tau_v A_v \frac{d s_v}{dt} & = &   s_v(t) \exp\left(-\frac{s_v(t)^2}{8a^2}\right)\left(\frac{\tau_v}{\tau A_u}A_v^2 - mA_u \right)  + \frac{\tau_v A_v}{\tau A_u}\int_{x} I^{ext}(x,t)u_1(x)dx. 
  \end{eqnarray}

To sum up, we get the original of Eqs.~(14-17).

\section*{Appendix C: travelling wave solution}\label{Appendix C}

We further analyze the condition for the network holding a continuously moving bump (travelling wave) as its stationary state. In such a state, the bump moves at a constant speed and its position is expressed as,
\begin{eqnarray}
    z(t)=v_{int}t,
\end{eqnarray}
where $v_{int}$ is the intrinsic speed of the network explained in Sec.3. In the travelling state, the bump height  and the discrepancy 
$d$ is a constant. With these conditions, Eqs.~(\ref{dynamics_u(t)_eqn},\ref{dynamics_v(t)_eqn}) are simplified to be,
\begin{eqnarray}
  \tau\left(A_u \frac{x-z}{2a^2}v_{int}\right)\mathcal{N}(z,2a) & = &
    (-A_u+\frac{\rho J_0}{\sqrt{2}}A_r)\mathcal{N}(z,2a) -A_v\mathcal{N}(z-s_v,2a),
    \label{dynamics_u(t)_eqn_travel}
    \\
    \tau_v\left(A_v \frac{x-z+s_v}{2a^2}v_{int}\right) \mathcal{N}(z-s_v,2a)
	 & = & -A_v\mathcal{N}(z-s_v,2a)+mA_u\mathcal{N}(z,2a).
    \label{dynamics_v(t)_eqn_travel}
\end{eqnarray}
Projecting both sides of Eq.~(\ref{dynamics_u(t)_eqn_travel}) onto the motion mode $u_0(x)$ (given by Eq.~(\ref{u0})), we obtain
  \begin{eqnarray}
	  Left & = & 0, \nonumber \\
	  Right & = & (-A_u+\frac{\rho J_0}{\sqrt{2}}A_r)\sqrt{2\pi}a-A_v\exp (-\frac{s_v^2}{8a^2})\sqrt{2\pi}a. \nonumber
  \end{eqnarray}
  Equating both sides, we have
  \begin{equation}
	  -A_u+\frac{\rho J_0}{\sqrt{2}} A_r-A_v\exp (-\frac{s_v^2}{8a^2})=0.
  \end{equation}
Similarly, projecting Eq.~(\ref{dynamics_u(t)_eqn_travel}) onto the motion mode $u_1(x)$ (given by Eq.~(\ref{u1})) and equating both sides, we obtain
  \begin{equation}
	  \tau A_u v_{int} =s_vA_v\exp(-\frac{s_v^2}{8a^2}).
  \end{equation}
Again, projecting both sides of Eq.~(\ref{dynamics_v(t)_eqn_travel}) onto the motion modes $u_0(x)$ and $u_1(x)$, respectively, and equating both sides, we obtain
  \begin{eqnarray}
	  \frac{s_v}{4a^2}\tau_v A_v\exp(-\frac{s_v^2}{8a^2})v_{int} & = & -A_v\exp(-\frac{s_v^2}{8a^2})+mA_u, 
	  \\
	  \tau_v(1-\frac{s_v^2}{4a^2})v_{int} & = & s_v.
  \end{eqnarray}
Summarize the four equations above, overall we have
\begin{eqnarray}
    A_u & = &  \frac{\rho J_0}{\sqrt{2}}A_r - A_v\exp \left(-\frac{s_v(t)^2}{8a^2}\right) ,
    \\
    \tau A_u v_{int} & = & s_v(t) A_v\exp\left(-\frac{s_v(t)^2}{8a^2}\right),
    \\
     A_v &=& mA_u \exp\left(-\frac{s_v(t)^2}{8a^2}\right),
    \\
    \tau_v A_v \frac{ds_v(t)}{dt} & = &  s_v(t) \exp\left(-\frac{s_v(t)^2}{8a^2}\right)\left(\frac{\tau_v}{\tau A_u}A_v^2 - mA_u \right),
\end{eqnarray}
Combining the above equations with Eq.~(\ref{dynamics_r_eqn}), we can analytically solve the five unknown parameters in travelling wave state:
  \begin{eqnarray}
  A_u & =& \frac{\rho J_0+\sqrt{\rho^2J_0^2-8\sqrt{2\pi}k\rho a(1+\sqrt{\frac{m\tau}{\tau_v}})^2}}
  {4\sqrt{\pi}k\rho a(1+\sqrt{\frac{m\tau}{\tau_v}})}, \\
  A_v & =& \frac{\rho J_0+\sqrt{\rho^2J_0^2-8\sqrt{2\pi}k\rho a(1+\sqrt{\frac{m\tau}{\tau_v}})^2}}{2\sqrt{2 \pi }k \rho ^2 a J_0}, \\
  A_r & =& \sqrt{\frac{m\tau}{\tau_v}}
  \exp\left[\frac{1-\sqrt{\frac{\tau}{m\tau_v}}}{2}\right]
  \frac{\rho J_0+\sqrt{\rho^2J_0^2-8\sqrt{2\pi}k\rho a(1+\sqrt{\frac{m\tau}{\tau_v}})^2}}{4\sqrt{\pi}k\rho a(1+\sqrt{\frac{m\tau}{\tau_v}})} , \\
  s_v &=& 2a\sqrt{1-\sqrt{\frac{\tau}{m\tau_v}}}, \\
  v_{int} &=& \frac{2a}{\tau_v}\sqrt{\frac{m\tau_v}{\tau}-\sqrt{\frac{m\tau_v}{\tau}}}.
  \end{eqnarray}

\bibliographystyle{unsrt}
\bibliography{NC_Template}	

\end{document}